\documentclass[aps,amssymb,amsmath,prl,reprint,noshowpacs]{revtex4-1}
\usepackage{times}
\usepackage{amssymb,amsmath,graphicx}
\usepackage[usenames]{color}
\usepackage{bbold}
\usepackage{siunitx}
\usepackage{braket}
\usepackage{comment}
\usepackage[T1]{fontenc}
\usepackage[utf8]{inputenc}
\usepackage[english]{babel}

\usepackage{hyperref}
\hypersetup{colorlinks=true, citecolor=blue, urlcolor=blue, linkcolor=blue}

\usepackage[normalem]{ulem}
\usepackage{lineno}

\newcommand{\Og}{\ensuremath{\Omega}}
\newcommand{\Ox}{\Omega_\mathrm{x}}
\newcommand{\Oy}{\Omega_\mathrm{y}}
\newcommand{\Oz}{\Omega_\mathrm{z}}

\newcommand{\Geff}{\gamma_\mathrm{eff}}
\newcommand{\Gmeas}{\Gamma_\mathrm{meas}}
\newcommand{\Gtot}{\Gamma_\mathrm{tot}}

\newcommand{\Gqba}{\Gamma_\mathrm{qba}}
\newcommand{\nbar}{\bar{n}}
\newcommand{\Sbar}{\bar{S}}
\newcommand{\zzpf}{z_\mathrm{zpf}}

\newcommand{\imu}{\text{\rm i}}
\newcommand{\diff}{\text{d}}
\usepackage{textcomp}

\newcommand{\commentOut}[1]{}
\usepackage{scalefnt}   

\newcommand{\affil}{Photonics Laboratory, ETH Zürich, CH-8093 Zürich, Switzerland}

\newcommand{\equalcontribution}{These authors contributed equally to this work.}

\begin{document}
\scalefont{1.05}
\title{Quantum control of a nanoparticle optically levitated in cryogenic free space}

\author{Felix Tebbenjohanns}
\altaffiliation{\equalcontribution}
\affiliation{\affil}
\author{M. Luisa Mattana}
\altaffiliation{\equalcontribution}
\affiliation{\affil}
\author{Massimiliano Rossi}
\altaffiliation{\equalcontribution}
\affiliation{\affil}
\author{Martin Frimmer}
\affiliation{\affil}
\author{Lukas Novotny}
\affiliation{\affil}
\homepage{http://www.photonics.ethz.ch}


\date\today

\maketitle

\textbf{ 
Tests of quantum mechanics on a macroscopic scale require extreme control over mechanical motion and its decoherence~\cite{Zurek2007,Chen2013,Arndt1999,Hornberger2012}.
Quantum control of mechanical motion has been achieved by engineering the radiation-pressure coupling between a micromechanical oscillator and the electromagnetic field in a resonator~\cite{Teufel2011,Chan2011,Qiu2020,Aspelmeyer2014}.
Furthermore, measurement-based feedback control relying on cavity-enhanced detection schemes has been used to cool micromechanical oscillators to their quantum ground states~\cite{Rossi2018}.
In contrast to mechanically tethered systems,
optically levitated nanoparticles are particularly promising candidates for matter-wave experiments with massive objects~\cite{Chang2010, Romero-Isart2010}, since their trapping potential is fully controllable.
In this work, we optically levitate a femto-gram dielectric particle in cryogenic free space, which suppresses thermal effects sufficiently to make the measurement backaction the dominant decoherence mechanism.
With an efficient quantum measurement, we exert quantum control over the dynamics of the particle.
We cool its center-of-mass motion by measurement-based feedback to an average occupancy of 0.65 motional quanta, corresponding to a state purity of 43\%. 
The absence of an optical resonator and its bandwidth limitations holds promise to transfer the full quantum control available for electromagnetic fields to a mechanical system. 
Together with the fact that the optical trapping potential is highly controllable, our experimental platform offers a route to investigating quantum mechanics at macroscopic scales~\cite{Romero-Isart2011,Leggett2002}.
}

\paragraph{Introduction.}
Mechanical oscillators with small dissipation have become indispensable tools for sensing and signal transduction~\cite{Braginskii1977,Verhagen2012,Andrews2014,Bagci2014,Mirhosseini2020}. In optomechanics, such oscillators are coupled to a light field to read out and control the mechanical motion at the fundamental limits set by quantum theory~\cite{Aspelmeyer2014}. A landmark feat in this context has been cavity-cooling of micromechanical oscillators to their quantum ground state of motion using dynamical backaction~\cite{Teufel2011,Chan2011}. 

The remarkable success of cavity optomechanics as a technology platform attracted the attention of a scientific community seeking to test the limitations of quantum theory at macroscopic scales~\cite{Cirac1998,Bose1999,Leggett1985,Leggett2002,Marshall2003}. A particularly exciting idea is to delocalize the wave function of a massive object over a distance larger than its physical size~\cite{Romero-Isart2011}. This regime is outside the scope of mechanically clamped oscillators and requires systems with largely tunable potentials, such as dielectric particles levitated in an optical trap~\cite{Romero-Isart2010,Chang2010}.  
The optical intensity distribution in a laser focus forms a controllable conservative potential for the particle's center-of-mass motion~\cite{Ashkin1977}.
A prerequisite for investigating macroscopic quantum effects is to prepare the particle in a quantum mechanically pure state, such as its motional ground state.
Subsequently, the trapping potential can be switched off~\cite{Hebestreit2018}, allowing for coherent evolution of the particle in the absence of decoherence generated by photon recoil heating~\cite{Purdy2013,Jain2016}. 
Furthermore, other sources of decoherence, such as collisions with gas molecules and recoil from blackbody photons, must be excluded~\cite{Kaltenbaek2012,Romero-Isart2011b}.
A cryogenic environment can provide both the required extreme high vacuum and the sufficiently low thermal population of the electromagnetic continuum.

Cavity-control of the center-of-mass motion of a levitated particle has made tremendous progress in recent years~\cite{Kiesel2013,Windey2019,Delic2019}, and ground-state cooling by dynamical back-action has recently been reported~\cite{Delic2020}.
An alternative approach to purify the particle's motional state relies on measurement-based feedback ~\cite{,Ashkin1977,Li2011,Gieseler2012, Iwasaki2019, Conangela2019, Tebbenjohanns2019}.
To operate this technique in the quantum regime requires performing a measurement whose quantum backaction represents the dominant disturbance of the system~\cite{Purdy2013,Jain2016}.
In addition, the result of this measurement needs to be recorded with sufficient efficiency, to compensate the measurement backaction by the feedback system~\cite{Mancini1998,Wilson2015,Rossi2018}.
Borrowing techniques developed for tethered optomechanical systems~\cite{Cohadon1999,Poggio2007,Wilson2015,Rossi2018}, levitated particles have been feedback-cooled to single-digit phonon occupation numbers~\cite{Kamba2020}, where first signatures of their motional ground state have been observed~\cite{Tebbenjohanns2020}. 
These studies suggest that ground-state cooling of mechanical motion without enhancing light-matter interaction with an optical resonator is possible with sufficiently high detection efficiency. Such a cavity-free optomechanical system would be unrestricted by the limitations regarding bandwidth, stability, and mode-matching associated with an optical resonator.

In this work, we optically levitate a nanoparticle in a cryogenic environment and feedback-cool its motion to the quantum ground state. Our feedback control relies on a {cavity-free} {optical measurement of the particle position that approaches the minimum of the Heisenberg relation} to within a factor of two. 

\paragraph{Experimental system.}

\begin{figure}[tb]\includegraphics{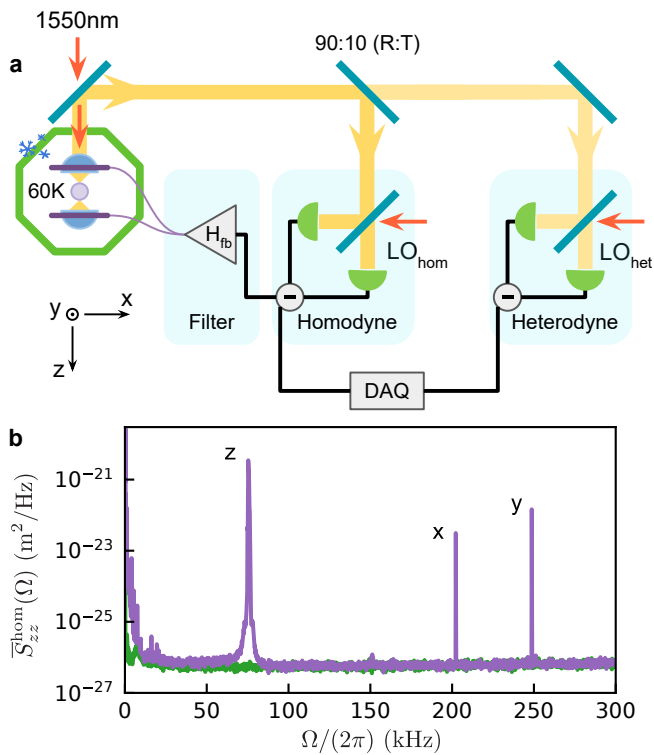}
\caption{{\bf Experimental setup.} {\bf (a)} An electrically charged silica nanoparticle is optically levitated in a cryogenic environment. The light scattered back by the particle is split between the heterodyne and the homodyne receivers.
The homodyne signal is filtered, and fed back as an electric force to the particle to cool its center-of-mass motion along the optical axis.
{\bf (b)} Power spectral density of the parametrically pre-cooled center-of-mass oscillation modes (purple) along the $z$, $x$, and $y$ axis (at 77~kHz, 202~kHz, and 249~kHz, respectively).
In green we plot the LO noise floor.
}
\label{fig:setup}
\end{figure}

In Fig.~\ref{fig:setup}a we show our experimental system. We generate a single-beam dipole trap by strongly focusing a laser ($P_t\sim 1.2$~W, wavelength $\lambda=1550$~nm, linearly polarized along the $x$ axis) with an aspheric trapping lens (numerical aperture 0.75).
A dipolar dielectric scatterer in the focal region experiences a three-dimensional confining potential, which is harmonic for small displacements from the focal center.  
In our experiments, we trap a single, electrically charged spherical silica nanoparticle (diameter 100~nm, mass $m\sim1$~fg).
The resonance frequency of the particle's center-of-mass motion along the optical axis $z$ is $\Oz/(2\pi)=77.6$~kHz (see Fig.~\ref{fig:setup}b).
The resonance frequencies in the focal plane are $\Ox/(2\pi)=202$~kHz along and $\Oy/(2\pi)=249$~kHz perpendicular to the axis of polarization.

To suppress heating due to collisions with gas molecules, we operate our optical trap inside a $4$ K cryostat.
On the holder of the trapping lens, we measure a temperature of $60$~K, which results from heating due to residual optical absorption (see Supplementary). The cryogenic environment reduces the thermal energy of the gas molecules, and simultaneously lowers the gas pressure by cryogenic pumping.
An ionization gauge located in the outer chamber (at $295$~K) of the cryostat reads a pressure of $3\times10^{-9}$~mbar, which we treat as an upper bound for the pressure at the location of the particle.
To stabilize the particle inside the trap and to avoid nonlinearities of the trapping potential, we pre-cool the particle's motion in the three dimensions using parametric feedback ~\cite{Gieseler2012}.
In the following, we focus our attention on the motion along the optical $z$ axis.
%

The detection of the particle's motion relies on the fact that its position is predominantly encoded in the phase of the light scattered back into the trapping lens~\cite{Tebbenjohanns2019Efficiency}.
This backscattered field is directed by an optical circulator to the detection setup, where $90$\% ($10$\%) of the signal is sent to a homodyne (heterodyne) receiver.
These receivers convert the phase of the optical field into an electrical signal. 
We use the homodyne measurement for feedback-control, and the heterodyne signal for an independent out-of-loop measurement of the particle's motion.
%

\paragraph{Feedback cooling to the ground state.}
\begin{figure*}[tb]
\includegraphics[width=\textwidth]{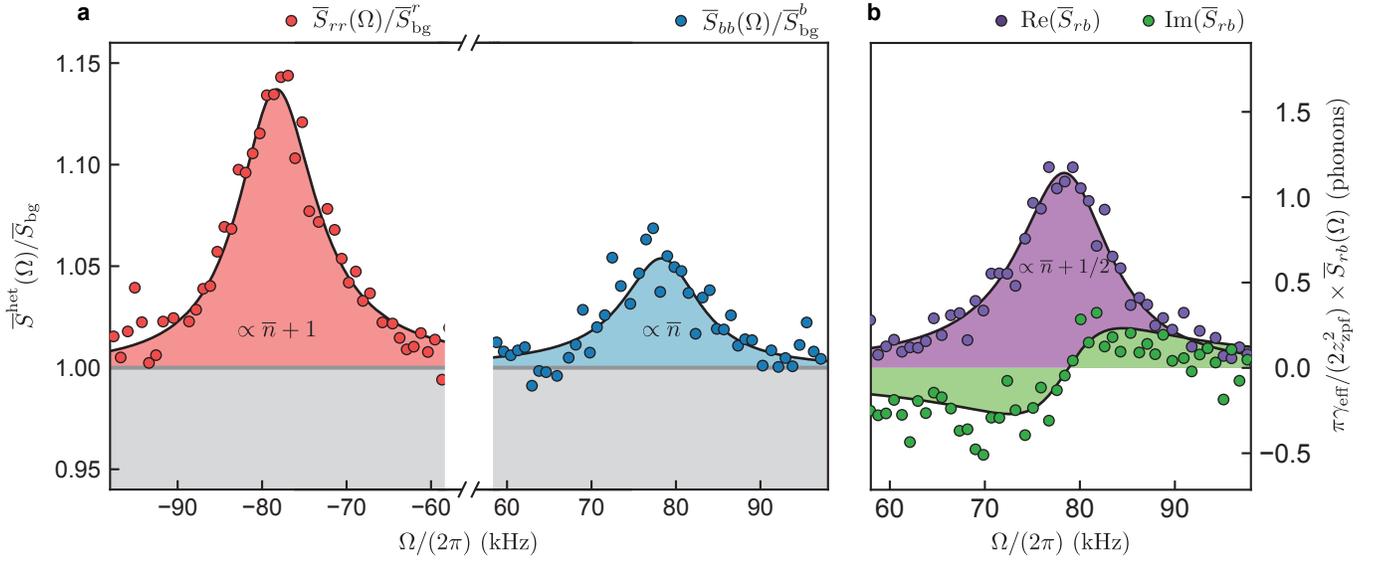}
\caption{{\bf Quantum ground state verification via out-of-loop measurements.} {\bf (a)} Stokes (red circles) and anti-Stokes (blue circles) sidebands measured by the out-of-loop heterodyne detector, at the largest electronic feedback gain. The black lines are fits to Eqs.~\eqref{eq:SrrSbb}, from which we extract the sideband powers. From their ratio, we extract a final occupation of $\nbar = 0.66\pm0.08$.
{\bf (b)} Real (purple circles) and imaginary (green circles) parts of the cross-power spectral density between the Stokes and anti-Stokes sideband, together with theoretical fits (black lines). We calibrate the vertical axis using the imaginary part, and we extract a final occupation of $\nbar = 0.64\pm0.09$ from the real part.
}
\label{fig:outofloop}
\end{figure*}
%
%
%
Our experimental platform is a cavity-free optomechanical system, performing a continuous measurement of the displacement of the particle \cite{Chang2010, Aspelmeyer2014}.
According to quantum theory, this measurement inevitably entails a backaction. For the levitated particle, this quantum backaction is associated with the radiation pressure shot noise arising from the quantization of the light field's linear momentum~\cite{Jain2016}.
Importantly, with a sufficiently efficient detection system in place (see Supplementary), it is possible to apply a feedback force to the particle that fully balances the effect of the backaction~\cite{Mancini1998,Cohadon1999,Rossi2018}.

We deploy a feedback method termed \emph{cold damping}~\cite{Mancini1998,Genes2008}. In this scheme, a viscous feedback force is derived from the measurement signal, increasing the dissipation while adding a minimum amount of fluctuations.
Our feedback circuit is a digital filter that electronically processes the homodyne signal in real-time. The filter mainly comprises a delay line to shift the phase of the frequencies near $\Og_z$ by $\pi/2$ (see Supplementary).
This procedure exploits the particle's harmonic motion to estimate the velocity from the measured displacement.
The filtered signal is applied as a voltage to a pair of electrodes located near the nanoparticle, actuating the feedback via the Coulomb force.

We now turn to the analysis of the particle's motional energy under feedback. Our first method to extract the phonon population of the particle relies on Raman sideband thermometry~\cite{Clerk2010, Safavi-Naeini2012, Tebbenjohanns2020}. To this end, we analyze the signal recorded on the heterodyne receiver (see Supplementary), which provides an out-of-loop measurement of the motion of the particle~\cite{Tebbenjohanns2019}. 
The power spectral density (PSD)~\footnote{We define our two-sided, symmetrized PSDs $\Sbar_{zz}(\Og)$ and our single-sided PSDs $\tilde{S}_{zz}(f)=4\pi \Sbar_{zz}(2\pi f)$ according to $\langle z^2 \rangle = \int_{-\infty}^\infty \diff \Og~ \Sbar_{zz}(\Og)= \int_0^\infty \diff f~\tilde{S}_{zz}(f)$} of both the red-shifted Stokes sideband $\Sbar_{rr}(\Og)$ and of the blue-shifted anti-Stokes sideband $\Sbar_{bb}(\Og)$ (Fig.~\ref{fig:outofloop}a) show a Lorentzian lineshape on top of a white noise floor.
Importantly, the total noise power in the two sidebands is visibly different. From this sideband asymmetry, we can extract the phonon population by fitting our data to the expressions
\begin{subequations}
\label{eq:SrrSbb}
\begin{align}
    \Sbar_{rr}(\Omega) &= \Sbar^r_\text{bg} + R |\chi_\text{eff}(\Og)|^2 (\nbar+1), \\
    \Sbar_{bb}(\Omega) &= \Sbar^b_\text{bg} + R |\chi_\text{eff}(\Og)|^2 \nbar,
\end{align}
\end{subequations}
with $\Sbar^{r,b}_\text{bg}$ the spectral background floor, $R=m\gamma_\text{eff}\hbar\Oz/\pi$ a scaling factor, $\chi_\text{eff}(\Og)=m^{-1}/(\Oz^2-\Og^2-\imu\Geff\Og)$ the effective mechanical susceptibility modified by the feedback, $\Geff$ the effective linewidth including the broadening due to feedback, and $\nbar$ the average phonon occupation of the mechanical state.

From the fit of our data (solid lines in Fig.~\ref{fig:outofloop}a), we extract a linewidth of $\Geff/(2\pi)=11.1$~kHz together with a residual occupation of $\nbar=0.66\pm0.08$, corresponding to a ground-state occupancy of $1/(\nbar+1)=60\%$.
The error is obtained by propagating the standard deviation (s.d.) of the fitted areas.
We note that the method of Raman thermometry does not rely on any calibration of the system. Instead, it is the zero-point energy of the oscillator which serves as the absolute scale all energies are measured against.

As a second method to infer the residual phonon population of the particle under feedback, we analyze the cross-correlations between the two measured sidebands~\cite{Purdy2017, Shkarin2019}. In Fig.~\ref{fig:outofloop}b, we show the real part of the measured cross correlation $\text{Re}(S_\text{rb})$ (purple) and its imaginary part $\text{Im}(S_\text{rb})$ (green). 
We fit the data to a theoretical model given by (see Supplementary)
\begin{equation}\label{eq:Srb}
\begin{split}
    \Sbar_{rb}(\Og) &= R |\chi_\text{eff}(\Og)|^2 \left(\nbar+\frac{1}{2}  + \frac{\imu}{2} \frac{\Og^2-\Oz^2}{\Geff\Oz}\right).
\end{split}
\end{equation}
Importantly, the imaginary part of the cross-correlation is independent of the phonon population $\nbar$. It arises purely from the zero-point fluctuations and can thus serve to calibrate the real part, from which we extract a phonon occupation of $\nbar=0.64\pm0.09$. 
The error is obtained from the propagation of the uncertainties (s.d.) in the fitted parameters.
This result is well in agreement with the value extracted from the sideband asymmetry.
%

\paragraph{Quantum measurement.}
\begin{figure*}[tb]
\includegraphics[width=\textwidth]{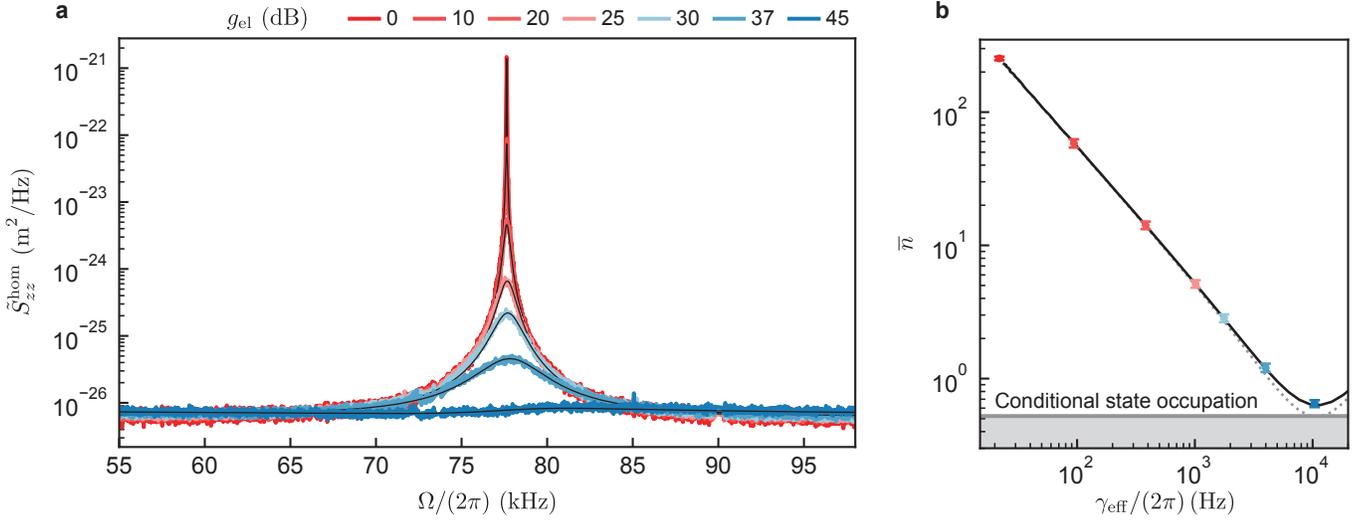}
\caption{{\bf In-loop analysis of the feedback system.} {\bf (a)} Single-sided displacement spectra measured by the in-loop homodyne detector, at different electronic gains $g_\text{el}$. We exclude three narrow spectral features from the analysis (see Supplementary). The black lines are fits to a theoretical model (see Supplementary). {\bf (b)} Mechanical occupations extracted from integrating the computed position and momentum spectra, which are based on parameters estimated from the in-loop spectra. The solid black (dotted grey) line is a theoretical model assuming an ideal delay filter (cold damping). The horizontal grey line corresponds to the occupation of the conditional state, stemming from the performed measurements. The error bars reflect the standard deviation (s.d.) in the fitted parameters, as well as the statistical error on the calibration method.}
\label{fig:inloop}
\end{figure*}
%
%
%
Efficient quantum measurement is a prerequisite for stabilizing the levitated nanoparticle in its quantum ground state via feedback.
In the following, we perform a detailed analysis of our measurement system.
To this end, we analyze the measurement record of our in-loop homodyne receiver and derive the measurement efficiency $\eta_\text{meas}$, that is, the amount of information gathered per disturbance incurred \cite{Wiseman2010}.
In Fig.~\ref{fig:inloop}a we show, in dark red, the homodyne spectrum acquired at the lowest feedback gain labelled by the set gain $g_\text{el}=0$~dB ($\Geff=2\pi\times 21.9$~Hz).
At such low gain, the measured fluctuations on resonance largely exceed the imprecision noise and the feedback solely leads to a broadening of the mechanical susceptibility.
In this regime, the imprecision noise fed back as a force does not play any role, and can be safely ignored.
Upon calibration via an out-of-loop energy measurement at a moderate gain (at $g_\text{el} = 25~$dB), we fit the observed spectrum to (see Supplementary)
\begin{eqnarray}
  \label{eq:disp-spec}
  \Sbar^\text{hom}_{zz}(\Og) = \Sbar_\text{imp} + |\chi_\text{eff}(\Og)|^2 \Sbar_{FF}^\text{tot},
\end{eqnarray}
where $\Sbar_{FF}^\text{tot}=\hbar^2\Gamma_\text{tot}/(2\pi\zzpf^2)$ is the total force noise  PSD, $\Sbar_\text{imp}= \zzpf^2/(8\pi \Gmeas)$ is the imprecision noise PSD,  and {$\zzpf^2=\hbar/(2m\Oz)$} denotes the zero-point fluctuations of the oscillator.
We note that these two spectral densities can be equivalently written in terms of a measurement rate $\Gmeas=\eta_d\Gqba$ (with $\Gqba$ the decoherence rate due to the quantum backaction, and $\eta_d$ the overall detection efficiency), and a total decoherence rate $\Gamma_\text{tot}=\Gqba + \Gamma_\text{exc}=\Geff (\nbar+1/2)$ (with $\Gamma_\text{exc}$ the decoherence rate in excess of quantum backaction).
From the fit, we extract a measurement rate of $\Gmeas/(2\pi)=(1.33\pm0.04)$~kHz 
and a total decoherence rate of $\Gamma_\text{tot}/(2\pi)=(5.5\pm0.3)$~kHz.
The measurement rate approaches the total decoherence rate, giving a measurement efficiency of $\eta_\text{meas}=\Gmeas/\Gamma_\text{tot}=0.24\pm0.02$, which is bounded by $\eta_\text{meas}\le1$ according to the Heisenberg measurement-disturbance relation~\cite{Braginsky1992, Clerk2010, Wiseman2010}.
%

Next, we characterize the role of the feedback gain in our system. To this end, we record homodyne spectra at increasing gain settings, as shown in Fig.~\ref{fig:inloop}a.
For small gain values, the feedback only increases the mechanical linewidth. For high gain values however, the spectra flatten and even dip below the imprecision noise, an effect known as {\it noise squashing} \cite{Cohadon1999}.
In this case, the feedback-induced correlations become dominant and increase the displacement fluctuations, rather than reducing them.
We fit each spectrum to a full in-loop model, where we independently characterize the transfer function of the electronic loop (see Supplementary).
Then, we use the results of the fits to compute the effective linewidths and the phonon occupations, shown in Fig.~\ref{fig:inloop}b.
At the highest gain, we estimate an occupation of $\nbar=(0.65\pm0.04)$, consistent with both other methods described above.
Based on the estimated measurement and total decoherence rates, we calculate a theoretical model for the occupations under a pure delay filter (black line in Fig.~\ref{fig:inloop}b).
For comparison, we show the theoretical results achievable under ideal cold damping \cite{Mancini1998} in the limit of $\Geff\ll\Oz$ (dotted grey line).
In this case, an induced linewidth of $\Geff$  corresponds to an occupation $\nbar=\Gtot/\Geff + \Geff/(16\Gmeas)-1/2$ \cite{Tebbenjohanns2019}, dependent only on the measurement and decoherence rates.
%

\paragraph{Discussion and outlook.}
In summary, we have achieved quantum control over the motion of a levitated nanosphere.
This control relies on the high reported measurement efficiency of $24$\%, comparable to what has been achieved with tethered micromechanical resonators~\cite{Rossi2018}, atomic systems~\cite{Sayrin2011}, and superconducting circuits~\cite{Vijay2012}.
As an example of measurement-based quantum control, we have experimentally stabilized the nanoparticle's motion in its quantum ground state via active feedback.
The prepared quantum state has a residual occupation of $\nbar=0.65$ phonons, corresponding to a purity of $1/(1+2\nbar)=43$\%.
Under optimal control, achievable by optimization of the feedback circuit, we expect to reach the same occupation as the conditional state \cite{Doherty1999b, Wiseman2010}, that is, $\nbar_\text{cond}\approx(1/\sqrt{\eta_\text{meas}}-1)/2=0.5$ (see Fig.~\ref{fig:inloop}b).
Our experiment approaches this limit to within $30$\%.
Notably, this is the first time that quantum control of mechanical degrees of freedom has been achieved without the use of an optical resonator.
In a study conducted in parallel to ours, similar results have been achieved with an optimal-control approach~\cite{Magrini2020}.
Our cavity-free platform allows overcoming the bistability in continuously operated optomechanical cavities, which limits the fastest achievable control time, $1/\Gqba$, to roughly the mechanical oscillation period $2\pi/\Og_z$ \cite{Aspelmeyer2014}.
The control time $1/\Gqba$ is inversely proportional to the particle's volume. When the excess decoherence is negligible, we expect to achieve $1/\Gqba\approx1/\Gtot=1\,\mu\text{s}$ for a 300-nm-diameter nanosphere, well below the measured period of  $2\pi/\Og_z=13\,\mu\text{s}$.
This opens the door for fast continuous and pulsed displacement measurement \cite{Meng2020, Vanner2011}.

Importantly, we conduct levitated-optomechanics experiments in a cryogenic environment for the first time.
This represents a milestone towards the generation of genuine macroscopic quantum states of a nanosphere, which would require extremely low levels of decoherence~\cite{Romero-Isart2011}.
On the one hand, cryogenic pumping can achieve extreme-high-vacuum in excess of $10^{-17}$~mbar \cite{Gabrielse1990}, suppressing decoherence due to gas collisions.
On the other hand, silica nanospheres quickly thermalize at the temperature of the surrounding cryogenic environment once the laser is switched off.
This drastically reduces the decoherence due to emission of blackbody photons.
For a trapping field intensity of $300~\text{mW/}\mu\text{m}^2$, the bulk heating rate due to optical absorption is estimated to be approximately $2$~K/ms \cite{Bateman2014}.
By switching on the optical field only for the needed duration of $1/\Gmeas\approx100~\mu\text{s}$ to stabilize the ground state~\cite{Doherty1999}, we can maintain the internal temperature of the nanosphere in equilibrium  with the surrounding cryogenic environment.
At the measured temperature of $60$~K and at a pressure of $10^{-12}$~mbar, well within the reach of state-of-the-art cryostats~\cite{Micke2019}, we estimate a coherent evolution time of around $50$~ms~\cite{Romero-Isart2011b}.
This would be sufficient to coherently expand the quantum wave function up to a size comparable with the nanosphere itself, opening the doors for exploring macroscopic quantum effects \cite{Kaltenbaek2016}.

\begin{acknowledgments}
\paragraph{Acknowledgments} This research was supported by the Swiss National Science Foundation (SNF) through the NCCR-QSIT program (Grant No.\ 51NF40-160591) and the the R'Equip program (Grant No. 206021-189605). 
We are grateful to F.~van der Laan for his contributions to the particle characterization procedure.
We thank O.~Wipfli and C.~Fischer for their suggestions in designing the cryogenic vacuum chamber,
J.~Piotrowski and D.~Windey for their advise with the trap assembly, 
and Y.~Li for her work on the control software.
We thank our colleagues P.~Back, E.~Bonvin, J.~Gao, A. Militaru, R. Reimann, J.~Vijayan, and J.~Zielinska for input and discussions. \\
\end{acknowledgments}

\bibliography{Literature}

\begin{thebibliography}{63}%
\makeatletter
\providecommand \@ifxundefined [1]{%
 \@ifx{#1\undefined}
}%
\providecommand \@ifnum [1]{%
 \ifnum #1\expandafter \@firstoftwo
 \else \expandafter \@secondoftwo
 \fi
}%
\providecommand \@ifx [1]{%
 \ifx #1\expandafter \@firstoftwo
 \else \expandafter \@secondoftwo
 \fi
}%
\providecommand \natexlab [1]{#1}%
\providecommand \enquote  [1]{``#1''}%
\providecommand \bibnamefont  [1]{#1}%
\providecommand \bibfnamefont [1]{#1}%
\providecommand \citenamefont [1]{#1}%
\providecommand \href@noop [0]{\@secondoftwo}%
\providecommand \href [0]{\begingroup \@sanitize@url \@href}%
\providecommand \@href[1]{\@@startlink{#1}\@@href}%
\providecommand \@@href[1]{\endgroup#1\@@endlink}%
\providecommand \@sanitize@url [0]{\catcode `\\12\catcode `\$12\catcode
  `\&12\catcode `\#12\catcode `\^12\catcode `\_12\catcode `\%12\relax}%
\providecommand \@@startlink[1]{}%
\providecommand \@@endlink[0]{}%
\providecommand \url  [0]{\begingroup\@sanitize@url \@url }%
\providecommand \@url [1]{\endgroup\@href {#1}{\urlprefix }}%
\providecommand \urlprefix  [0]{URL }%
\providecommand \Eprint [0]{\href }%
\providecommand \doibase [0]{http://dx.doi.org/}%
\providecommand \selectlanguage [0]{\@gobble}%
\providecommand \bibinfo  [0]{\@secondoftwo}%
\providecommand \bibfield  [0]{\@secondoftwo}%
\providecommand \translation [1]{[#1]}%
\providecommand \BibitemOpen [0]{}%
\providecommand \bibitemStop [0]{}%
\providecommand \bibitemNoStop [0]{.\EOS\space}%
\providecommand \EOS [0]{\spacefactor3000\relax}%
\providecommand \BibitemShut  [1]{\csname bibitem#1\endcsname}%
\let\auto@bib@innerbib\@empty
\bibitem [{\citenamefont {Zurek}(2007)}]{Zurek2007}%
  \BibitemOpen
  \bibfield  {author} {\bibinfo {author} {\bibfnamefont {W.~H.}\ \bibnamefont
  {Zurek}},\ }\enquote {\bibinfo {title} {Decoherence and the transition from
  quantum to classical --- revisited},}\ in\ \href {\doibase
  10.1007/978-3-7643-7808-0_1} {\emph {\bibinfo {booktitle} {Quantum
  Decoherence: Poincar{\'e} Seminar 2005}}},\ \bibinfo {editor} {edited by\
  \bibinfo {editor} {\bibfnamefont {B.}~\bibnamefont {Duplantier}}, \bibinfo
  {editor} {\bibfnamefont {J.-M.}\ \bibnamefont {Raimond}}, \ and\ \bibinfo
  {editor} {\bibfnamefont {V.}~\bibnamefont {Rivasseau}}}\ (\bibinfo
  {publisher} {Birkh{\"a}user Basel},\ \bibinfo {address} {Basel},\ \bibinfo
  {year} {2007})\ pp.\ \bibinfo {pages} {1--31}\BibitemShut {NoStop}%
\bibitem [{\citenamefont {Chen}(2013)}]{Chen2013}%
  \BibitemOpen
  \bibfield  {author} {\bibinfo {author} {\bibfnamefont {Y.}~\bibnamefont
  {Chen}},\ }\href {\doibase 10.1088/0953-4075/46/10/104001} {\bibfield
  {journal} {\bibinfo  {journal} {Journal of Physics B: Atomic, Molecular and
  Optical Physics}\ }\textbf {\bibinfo {volume} {46}},\ \bibinfo {pages}
  {104001} (\bibinfo {year} {2013})}\BibitemShut {NoStop}%
\bibitem [{\citenamefont {Arndt}\ \emph {et~al.}(1999)\citenamefont {Arndt},
  \citenamefont {Nairz}, \citenamefont {Vos-Andreae}, \citenamefont {Keller},
  \citenamefont {Zouw},\ and\ \citenamefont {Zeilinger}}]{Arndt1999}%
  \BibitemOpen
  \bibfield  {author} {\bibinfo {author} {\bibfnamefont {M.}~\bibnamefont
  {Arndt}}, \bibinfo {author} {\bibfnamefont {O.}~\bibnamefont {Nairz}},
  \bibinfo {author} {\bibfnamefont {J.}~\bibnamefont {Vos-Andreae}}, \bibinfo
  {author} {\bibfnamefont {C.}~\bibnamefont {Keller}}, \bibinfo {author}
  {\bibfnamefont {G.~V.~D.}\ \bibnamefont {Zouw}}, \ and\ \bibinfo {author}
  {\bibfnamefont {A.}~\bibnamefont {Zeilinger}},\ }\href {\doibase
  10.1038/44348} {\bibfield  {journal} {\bibinfo  {journal} {Nature}\ }\textbf
  {\bibinfo {volume} {401}},\ \bibinfo {pages} {680–682} (\bibinfo {year}
  {1999})}\BibitemShut {NoStop}%
\bibitem [{\citenamefont {Hornberger}\ \emph {et~al.}(2012)\citenamefont
  {Hornberger}, \citenamefont {Gerlich}, \citenamefont {Haslinger},
  \citenamefont {Nimmrichter},\ and\ \citenamefont {Arndt}}]{Hornberger2012}%
  \BibitemOpen
  \bibfield  {author} {\bibinfo {author} {\bibfnamefont {K.}~\bibnamefont
  {Hornberger}}, \bibinfo {author} {\bibfnamefont {S.}~\bibnamefont {Gerlich}},
  \bibinfo {author} {\bibfnamefont {P.}~\bibnamefont {Haslinger}}, \bibinfo
  {author} {\bibfnamefont {S.}~\bibnamefont {Nimmrichter}}, \ and\ \bibinfo
  {author} {\bibfnamefont {M.}~\bibnamefont {Arndt}},\ }\href {\doibase
  10.1103/RevModPhys.84.157} {\bibfield  {journal} {\bibinfo  {journal} {Rev.
  Mod. Phys.}\ }\textbf {\bibinfo {volume} {84}},\ \bibinfo {pages} {157}
  (\bibinfo {year} {2012})}\BibitemShut {NoStop}%
\bibitem [{\citenamefont {Teufel}\ \emph {et~al.}(2011)\citenamefont {Teufel},
  \citenamefont {Donner}, \citenamefont {Li}, \citenamefont {Harlow},
  \citenamefont {Allman}, \citenamefont {Cicak}, \citenamefont {Sirois},
  \citenamefont {Whittaker}, \citenamefont {Lehnert},\ and\ \citenamefont
  {Simmonds}}]{Teufel2011}%
  \BibitemOpen
  \bibfield  {author} {\bibinfo {author} {\bibfnamefont {J.~D.}\ \bibnamefont
  {Teufel}}, \bibinfo {author} {\bibfnamefont {T.}~\bibnamefont {Donner}},
  \bibinfo {author} {\bibfnamefont {D.}~\bibnamefont {Li}}, \bibinfo {author}
  {\bibfnamefont {J.~W.}\ \bibnamefont {Harlow}}, \bibinfo {author}
  {\bibfnamefont {M.~S.}\ \bibnamefont {Allman}}, \bibinfo {author}
  {\bibfnamefont {K.}~\bibnamefont {Cicak}}, \bibinfo {author} {\bibfnamefont
  {A.~J.}\ \bibnamefont {Sirois}}, \bibinfo {author} {\bibfnamefont {J.~D.}\
  \bibnamefont {Whittaker}}, \bibinfo {author} {\bibfnamefont {K.~W.}\
  \bibnamefont {Lehnert}}, \ and\ \bibinfo {author} {\bibfnamefont {R.~W.}\
  \bibnamefont {Simmonds}},\ }\href {\doibase 10.1038/nature10261} {\bibfield
  {journal} {\bibinfo  {journal} {Nature}\ }\textbf {\bibinfo {volume} {475}},\
  \bibinfo {pages} {359} (\bibinfo {year} {2011})}\BibitemShut {NoStop}%
\bibitem [{\citenamefont {Chan}\ \emph {et~al.}(2011)\citenamefont {Chan},
  \citenamefont {Alegre}, \citenamefont {Safavi-Naeini}, \citenamefont {Hill},
  \citenamefont {Krause}, \citenamefont {Gr{\"o}blacher}, \citenamefont
  {Aspelmeyer},\ and\ \citenamefont {Painter}}]{Chan2011}%
  \BibitemOpen
  \bibfield  {author} {\bibinfo {author} {\bibfnamefont {J.}~\bibnamefont
  {Chan}}, \bibinfo {author} {\bibfnamefont {T.~P.~M.}\ \bibnamefont {Alegre}},
  \bibinfo {author} {\bibfnamefont {A.~H.}\ \bibnamefont {Safavi-Naeini}},
  \bibinfo {author} {\bibfnamefont {J.~T.}\ \bibnamefont {Hill}}, \bibinfo
  {author} {\bibfnamefont {A.}~\bibnamefont {Krause}}, \bibinfo {author}
  {\bibfnamefont {S.}~\bibnamefont {Gr{\"o}blacher}}, \bibinfo {author}
  {\bibfnamefont {M.}~\bibnamefont {Aspelmeyer}}, \ and\ \bibinfo {author}
  {\bibfnamefont {O.}~\bibnamefont {Painter}},\ }\href {\doibase
  10.1038/nature10461} {\bibfield  {journal} {\bibinfo  {journal} {Nature}\
  }\textbf {\bibinfo {volume} {478}},\ \bibinfo {pages} {89–92} (\bibinfo
  {year} {2011})}\BibitemShut {NoStop}%
\bibitem [{\citenamefont {Qiu}\ \emph {et~al.}(2020)\citenamefont {Qiu},
  \citenamefont {Shomroni}, \citenamefont {Seidler},\ and\ \citenamefont
  {Kippenberg}}]{Qiu2020}%
  \BibitemOpen
  \bibfield  {author} {\bibinfo {author} {\bibfnamefont {L.}~\bibnamefont
  {Qiu}}, \bibinfo {author} {\bibfnamefont {I.}~\bibnamefont {Shomroni}},
  \bibinfo {author} {\bibfnamefont {P.}~\bibnamefont {Seidler}}, \ and\
  \bibinfo {author} {\bibfnamefont {T.~J.}\ \bibnamefont {Kippenberg}},\ }\href
  {\doibase 10.1103/PhysRevLett.124.173601} {\bibfield  {journal} {\bibinfo
  {journal} {Phys. Rev. Lett.}\ }\textbf {\bibinfo {volume} {124}},\ \bibinfo
  {pages} {173601} (\bibinfo {year} {2020})}\BibitemShut {NoStop}%
\bibitem [{\citenamefont {Aspelmeyer}\ \emph {et~al.}(2014)\citenamefont
  {Aspelmeyer}, \citenamefont {Kippenberg},\ and\ \citenamefont
  {Marquardt}}]{Aspelmeyer2014}%
  \BibitemOpen
  \bibfield  {author} {\bibinfo {author} {\bibfnamefont {M.}~\bibnamefont
  {Aspelmeyer}}, \bibinfo {author} {\bibfnamefont {T.~J.}\ \bibnamefont
  {Kippenberg}}, \ and\ \bibinfo {author} {\bibfnamefont {F.}~\bibnamefont
  {Marquardt}},\ }\href {\doibase 10.1103/RevModPhys.86.1391} {\bibfield
  {journal} {\bibinfo  {journal} {Rev. Mod. Phys.}\ }\textbf {\bibinfo {volume}
  {86}},\ \bibinfo {pages} {1391} (\bibinfo {year} {2014})}\BibitemShut
  {NoStop}%
\bibitem [{\citenamefont {Rossi}\ \emph {et~al.}(2018)\citenamefont {Rossi},
  \citenamefont {Mason}, \citenamefont {Chen}, \citenamefont {Tsaturyan},\ and\
  \citenamefont {Schliesser}}]{Rossi2018}%
  \BibitemOpen
  \bibfield  {author} {\bibinfo {author} {\bibfnamefont {M.}~\bibnamefont
  {Rossi}}, \bibinfo {author} {\bibfnamefont {D.}~\bibnamefont {Mason}},
  \bibinfo {author} {\bibfnamefont {J.}~\bibnamefont {Chen}}, \bibinfo {author}
  {\bibfnamefont {Y.}~\bibnamefont {Tsaturyan}}, \ and\ \bibinfo {author}
  {\bibfnamefont {A.}~\bibnamefont {Schliesser}},\ }\href {\doibase
  10.1038/s41586-018-0643-8} {\bibfield  {journal} {\bibinfo  {journal}
  {Nature}\ }\textbf {\bibinfo {volume} {563}},\ \bibinfo {pages} {53–58}
  (\bibinfo {year} {2018})}\BibitemShut {NoStop}%
\bibitem [{\citenamefont {Chang}\ \emph {et~al.}(2010)\citenamefont {Chang},
  \citenamefont {Regal}, \citenamefont {Papp}, \citenamefont {Wilson},
  \citenamefont {Ye}, \citenamefont {Painter}, \citenamefont {Kimble},\ and\
  \citenamefont {Zoller}}]{Chang2010}%
  \BibitemOpen
  \bibfield  {author} {\bibinfo {author} {\bibfnamefont {D.~E.}\ \bibnamefont
  {Chang}}, \bibinfo {author} {\bibfnamefont {C.~A.}\ \bibnamefont {Regal}},
  \bibinfo {author} {\bibfnamefont {S.~B.}\ \bibnamefont {Papp}}, \bibinfo
  {author} {\bibfnamefont {D.~J.}\ \bibnamefont {Wilson}}, \bibinfo {author}
  {\bibfnamefont {J.}~\bibnamefont {Ye}}, \bibinfo {author} {\bibfnamefont
  {O.}~\bibnamefont {Painter}}, \bibinfo {author} {\bibfnamefont {H.~J.}\
  \bibnamefont {Kimble}}, \ and\ \bibinfo {author} {\bibfnamefont
  {P.}~\bibnamefont {Zoller}},\ }\href {\doibase 10.1073/pnas.0912969107}
  {\bibfield  {journal} {\bibinfo  {journal} {Proc. Natl. Acad. Sci. USA}\
  }\textbf {\bibinfo {volume} {107}},\ \bibinfo {pages} {1005} (\bibinfo {year}
  {2010})}\BibitemShut {NoStop}%
\bibitem [{\citenamefont {Romero-Isart}\ \emph {et~al.}(2010)\citenamefont
  {Romero-Isart}, \citenamefont {Juan}, \citenamefont {Quidant},\ and\
  \citenamefont {Cirac}}]{Romero-Isart2010}%
  \BibitemOpen
  \bibfield  {author} {\bibinfo {author} {\bibfnamefont {O.}~\bibnamefont
  {Romero-Isart}}, \bibinfo {author} {\bibfnamefont {M.~L.}\ \bibnamefont
  {Juan}}, \bibinfo {author} {\bibfnamefont {R.}~\bibnamefont {Quidant}}, \
  and\ \bibinfo {author} {\bibfnamefont {J.~I.}\ \bibnamefont {Cirac}},\ }\href
  {\doibase 10.1088/1367-2630/12/3/033015} {\bibfield  {journal} {\bibinfo
  {journal} {New Journal of Physics}\ }\textbf {\bibinfo {volume} {12}},\
  \bibinfo {pages} {033015} (\bibinfo {year} {2010})}\BibitemShut {NoStop}%
\bibitem [{\citenamefont {Romero-Isart}\ \emph {et~al.}(2011)\citenamefont
  {Romero-Isart}, \citenamefont {Pflanzer}, \citenamefont {Blaser},
  \citenamefont {Kaltenbaek}, \citenamefont {Kiesel}, \citenamefont
  {Aspelmeyer},\ and\ \citenamefont {Cirac}}]{Romero-Isart2011}%
  \BibitemOpen
  \bibfield  {author} {\bibinfo {author} {\bibfnamefont {O.}~\bibnamefont
  {Romero-Isart}}, \bibinfo {author} {\bibfnamefont {A.~C.}\ \bibnamefont
  {Pflanzer}}, \bibinfo {author} {\bibfnamefont {F.}~\bibnamefont {Blaser}},
  \bibinfo {author} {\bibfnamefont {R.}~\bibnamefont {Kaltenbaek}}, \bibinfo
  {author} {\bibfnamefont {N.}~\bibnamefont {Kiesel}}, \bibinfo {author}
  {\bibfnamefont {M.}~\bibnamefont {Aspelmeyer}}, \ and\ \bibinfo {author}
  {\bibfnamefont {J.~I.}\ \bibnamefont {Cirac}},\ }\href {\doibase
  10.1103/PhysRevLett.107.020405} {\bibfield  {journal} {\bibinfo  {journal}
  {Phys. Rev. Lett.}\ }\textbf {\bibinfo {volume} {107}},\ \bibinfo {pages}
  {020405} (\bibinfo {year} {2011})}\BibitemShut {NoStop}%
\bibitem [{\citenamefont {Leggett}(2002)}]{Leggett2002}%
  \BibitemOpen
  \bibfield  {author} {\bibinfo {author} {\bibfnamefont {A.~J.}\ \bibnamefont
  {Leggett}},\ }\href {\doibase 10.1088/0953-8984/14/15/201} {\bibfield
  {journal} {\bibinfo  {journal} {Journal of Physics: Condensed Matter}\
  }\textbf {\bibinfo {volume} {14}},\ \bibinfo {pages} {R415} (\bibinfo {year}
  {2002})}\BibitemShut {NoStop}%
\bibitem [{\citenamefont {Braginskii}\ and\ \citenamefont
  {Manukin}(1977)}]{Braginskii1977}%
  \BibitemOpen
  \bibfield  {author} {\bibinfo {author} {\bibfnamefont {V.~B.}\ \bibnamefont
  {Braginskii}}\ and\ \bibinfo {author} {\bibfnamefont {A.~B.}\ \bibnamefont
  {Manukin}},\ }\href@noop {} {{\selectlanguage {English}\emph {\bibinfo
  {title} {Measurement of weak forces in physics experiments / V. B. Braginsky
  and A. B. Manukin ; edited by David H. Douglass}}}}\ (\bibinfo  {publisher}
  {University of Chicago Press Chicago},\ \bibinfo {year} {1977})\BibitemShut
  {NoStop}%
\bibitem [{\citenamefont {Verhagen}\ \emph {et~al.}(2012)\citenamefont
  {Verhagen}, \citenamefont {Deléglise}, \citenamefont {Weis}, \citenamefont
  {Schliesser},\ and\ \citenamefont {Kippenberg}}]{Verhagen2012}%
  \BibitemOpen
  \bibfield  {author} {\bibinfo {author} {\bibfnamefont {E.}~\bibnamefont
  {Verhagen}}, \bibinfo {author} {\bibfnamefont {S.}~\bibnamefont
  {Deléglise}}, \bibinfo {author} {\bibfnamefont {S.}~\bibnamefont {Weis}},
  \bibinfo {author} {\bibfnamefont {A.}~\bibnamefont {Schliesser}}, \ and\
  \bibinfo {author} {\bibfnamefont {T.~J.}\ \bibnamefont {Kippenberg}},\ }\href
  {\doibase 10.1038/nature10787} {\bibfield  {journal} {\bibinfo  {journal}
  {Nature}\ }\textbf {\bibinfo {volume} {482}},\ \bibinfo {pages} {63–67}
  (\bibinfo {year} {2012})}\BibitemShut {NoStop}%
\bibitem [{\citenamefont {Andrews}\ \emph {et~al.}(2014)\citenamefont
  {Andrews}, \citenamefont {Peterson}, \citenamefont {Purdy}, \citenamefont
  {Cicak}, \citenamefont {Simmonds}, \citenamefont {Regal},\ and\ \citenamefont
  {Lehnert}}]{Andrews2014}%
  \BibitemOpen
  \bibfield  {author} {\bibinfo {author} {\bibfnamefont {R.~W.}\ \bibnamefont
  {Andrews}}, \bibinfo {author} {\bibfnamefont {R.~W.}\ \bibnamefont
  {Peterson}}, \bibinfo {author} {\bibfnamefont {T.~P.}\ \bibnamefont {Purdy}},
  \bibinfo {author} {\bibfnamefont {K.}~\bibnamefont {Cicak}}, \bibinfo
  {author} {\bibfnamefont {R.~W.}\ \bibnamefont {Simmonds}}, \bibinfo {author}
  {\bibfnamefont {C.~A.}\ \bibnamefont {Regal}}, \ and\ \bibinfo {author}
  {\bibfnamefont {K.~W.}\ \bibnamefont {Lehnert}},\ }\href {\doibase
  10.1038/nphys2911} {\bibfield  {journal} {\bibinfo  {journal} {Nature
  Physics}\ }\textbf {\bibinfo {volume} {10}},\ \bibinfo {pages} {321}
  (\bibinfo {year} {2014})}\BibitemShut {NoStop}%
\bibitem [{\citenamefont {Bagci}\ \emph {et~al.}(2014)\citenamefont {Bagci},
  \citenamefont {Simonsen}, \citenamefont {Schmid}, \citenamefont {Villanueva},
  \citenamefont {Zeuthen}, \citenamefont {Appel}, \citenamefont {Taylor},
  \citenamefont {S{\o}rensen}, \citenamefont {Usami}, \citenamefont
  {Schliesser},\ and\ \citenamefont {Polzik}}]{Bagci2014}%
  \BibitemOpen
  \bibfield  {author} {\bibinfo {author} {\bibfnamefont {T.}~\bibnamefont
  {Bagci}}, \bibinfo {author} {\bibfnamefont {A.}~\bibnamefont {Simonsen}},
  \bibinfo {author} {\bibfnamefont {S.}~\bibnamefont {Schmid}}, \bibinfo
  {author} {\bibfnamefont {L.~G.}\ \bibnamefont {Villanueva}}, \bibinfo
  {author} {\bibfnamefont {E.}~\bibnamefont {Zeuthen}}, \bibinfo {author}
  {\bibfnamefont {J.}~\bibnamefont {Appel}}, \bibinfo {author} {\bibfnamefont
  {J.~M.}\ \bibnamefont {Taylor}}, \bibinfo {author} {\bibfnamefont
  {A.}~\bibnamefont {S{\o}rensen}}, \bibinfo {author} {\bibfnamefont
  {K.}~\bibnamefont {Usami}}, \bibinfo {author} {\bibfnamefont
  {A.}~\bibnamefont {Schliesser}}, \ and\ \bibinfo {author} {\bibfnamefont
  {E.~S.}\ \bibnamefont {Polzik}},\ }\href {\doibase 10.1038/nature13029}
  {\bibfield  {journal} {\bibinfo  {journal} {Nature}\ }\textbf {\bibinfo
  {volume} {507}},\ \bibinfo {pages} {81} (\bibinfo {year} {2014})}\BibitemShut
  {NoStop}%
\bibitem [{\citenamefont {Mirhosseini}\ \emph {et~al.}(2020)\citenamefont
  {Mirhosseini}, \citenamefont {Sipahigil}, \citenamefont {Kalaee},\ and\
  \citenamefont {Painter}}]{Mirhosseini2020}%
  \BibitemOpen
  \bibfield  {author} {\bibinfo {author} {\bibfnamefont {M.}~\bibnamefont
  {Mirhosseini}}, \bibinfo {author} {\bibfnamefont {A.}~\bibnamefont
  {Sipahigil}}, \bibinfo {author} {\bibfnamefont {M.}~\bibnamefont {Kalaee}}, \
  and\ \bibinfo {author} {\bibfnamefont {O.}~\bibnamefont {Painter}},\ }\href
  {\doibase 10.1038/s41586-020-3038-6} {\bibfield  {journal} {\bibinfo
  {journal} {Nature}\ }\textbf {\bibinfo {volume} {588}},\ \bibinfo {pages}
  {599} (\bibinfo {year} {2020})}\BibitemShut {NoStop}%
\bibitem [{\citenamefont {Cirac}\ \emph {et~al.}(1998)\citenamefont {Cirac},
  \citenamefont {Lewenstein}, \citenamefont {M\o{}lmer},\ and\ \citenamefont
  {Zoller}}]{Cirac1998}%
  \BibitemOpen
  \bibfield  {author} {\bibinfo {author} {\bibfnamefont {J.~I.}\ \bibnamefont
  {Cirac}}, \bibinfo {author} {\bibfnamefont {M.}~\bibnamefont {Lewenstein}},
  \bibinfo {author} {\bibfnamefont {K.}~\bibnamefont {M\o{}lmer}}, \ and\
  \bibinfo {author} {\bibfnamefont {P.}~\bibnamefont {Zoller}},\ }\href
  {\doibase 10.1103/PhysRevA.57.1208} {\bibfield  {journal} {\bibinfo
  {journal} {Phys. Rev. A}\ }\textbf {\bibinfo {volume} {57}},\ \bibinfo
  {pages} {1208} (\bibinfo {year} {1998})}\BibitemShut {NoStop}%
\bibitem [{\citenamefont {Bose}\ \emph {et~al.}(1999)\citenamefont {Bose},
  \citenamefont {Jacobs},\ and\ \citenamefont {Knight}}]{Bose1999}%
  \BibitemOpen
  \bibfield  {author} {\bibinfo {author} {\bibfnamefont {S.}~\bibnamefont
  {Bose}}, \bibinfo {author} {\bibfnamefont {K.}~\bibnamefont {Jacobs}}, \ and\
  \bibinfo {author} {\bibfnamefont {P.~L.}\ \bibnamefont {Knight}},\ }\href
  {\doibase 10.1103/PhysRevA.59.3204} {\bibfield  {journal} {\bibinfo
  {journal} {Phys. Rev. A}\ }\textbf {\bibinfo {volume} {59}},\ \bibinfo
  {pages} {3204} (\bibinfo {year} {1999})}\BibitemShut {NoStop}%
\bibitem [{\citenamefont {Leggett}\ and\ \citenamefont
  {Garg}(1985)}]{Leggett1985}%
  \BibitemOpen
  \bibfield  {author} {\bibinfo {author} {\bibfnamefont {A.~J.}\ \bibnamefont
  {Leggett}}\ and\ \bibinfo {author} {\bibfnamefont {A.}~\bibnamefont {Garg}},\
  }\href {\doibase 10.1103/PhysRevLett.54.857} {\bibfield  {journal} {\bibinfo
  {journal} {Phys. Rev. Lett.}\ }\textbf {\bibinfo {volume} {54}},\ \bibinfo
  {pages} {857} (\bibinfo {year} {1985})}\BibitemShut {NoStop}%
\bibitem [{\citenamefont {Marshall}\ \emph {et~al.}(2003)\citenamefont
  {Marshall}, \citenamefont {Simon}, \citenamefont {Penrose},\ and\
  \citenamefont {Bouwmeester}}]{Marshall2003}%
  \BibitemOpen
  \bibfield  {author} {\bibinfo {author} {\bibfnamefont {W.}~\bibnamefont
  {Marshall}}, \bibinfo {author} {\bibfnamefont {C.}~\bibnamefont {Simon}},
  \bibinfo {author} {\bibfnamefont {R.}~\bibnamefont {Penrose}}, \ and\
  \bibinfo {author} {\bibfnamefont {D.}~\bibnamefont {Bouwmeester}},\ }\href
  {\doibase 10.1103/PhysRevLett.91.130401} {\bibfield  {journal} {\bibinfo
  {journal} {Phys. Rev. Lett.}\ }\textbf {\bibinfo {volume} {91}},\ \bibinfo
  {pages} {130401} (\bibinfo {year} {2003})}\BibitemShut {NoStop}%
\bibitem [{\citenamefont {Ashkin}\ and\ \citenamefont
  {Dziedzic}(1977)}]{Ashkin1977}%
  \BibitemOpen
  \bibfield  {author} {\bibinfo {author} {\bibfnamefont {A.}~\bibnamefont
  {Ashkin}}\ and\ \bibinfo {author} {\bibfnamefont {J.~M.}\ \bibnamefont
  {Dziedzic}},\ }\href@noop {} {\bibfield  {journal} {\bibinfo  {journal}
  {Appl. Phys. Lett.}\ }\textbf {\bibinfo {volume} {30}},\ \bibinfo {pages}
  {202} (\bibinfo {year} {1977})}\BibitemShut {NoStop}%
\bibitem [{\citenamefont {Hebestreit}\ \emph {et~al.}(2018)\citenamefont
  {Hebestreit}, \citenamefont {Frimmer}, \citenamefont {Reimann}, \citenamefont
  {Dellago}, \citenamefont {Ricci},\ and\ \citenamefont
  {Novotny}}]{Hebestreit2018}%
  \BibitemOpen
  \bibfield  {author} {\bibinfo {author} {\bibfnamefont {E.}~\bibnamefont
  {Hebestreit}}, \bibinfo {author} {\bibfnamefont {M.}~\bibnamefont {Frimmer}},
  \bibinfo {author} {\bibfnamefont {R.}~\bibnamefont {Reimann}}, \bibinfo
  {author} {\bibfnamefont {C.}~\bibnamefont {Dellago}}, \bibinfo {author}
  {\bibfnamefont {F.}~\bibnamefont {Ricci}}, \ and\ \bibinfo {author}
  {\bibfnamefont {L.}~\bibnamefont {Novotny}},\ }\href {\doibase
  10.1063/1.5017119} {\bibfield  {journal} {\bibinfo  {journal} {Rev. Sci.
  Instr.}\ }\textbf {\bibinfo {volume} {89}},\ \bibinfo {pages} {033111}
  (\bibinfo {year} {2018})}\BibitemShut {NoStop}%
\bibitem [{\citenamefont {Purdy}\ \emph {et~al.}(2013)\citenamefont {Purdy},
  \citenamefont {Peterson},\ and\ \citenamefont {Regal}}]{Purdy2013}%
  \BibitemOpen
  \bibfield  {author} {\bibinfo {author} {\bibfnamefont {T.~P.}\ \bibnamefont
  {Purdy}}, \bibinfo {author} {\bibfnamefont {R.~W.}\ \bibnamefont {Peterson}},
  \ and\ \bibinfo {author} {\bibfnamefont {C.~A.}\ \bibnamefont {Regal}},\
  }\href {\doibase 10.1126/science.1231282} {\bibfield  {journal} {\bibinfo
  {journal} {Science}\ }\textbf {\bibinfo {volume} {339}},\ \bibinfo {pages}
  {801} (\bibinfo {year} {2013})}\BibitemShut {NoStop}%
\bibitem [{\citenamefont {Jain}\ \emph {et~al.}(2016)\citenamefont {Jain},
  \citenamefont {Gieseler}, \citenamefont {Moritz}, \citenamefont {Dellago},
  \citenamefont {Quidant},\ and\ \citenamefont {Novotny}}]{Jain2016}%
  \BibitemOpen
  \bibfield  {author} {\bibinfo {author} {\bibfnamefont {V.}~\bibnamefont
  {Jain}}, \bibinfo {author} {\bibfnamefont {J.}~\bibnamefont {Gieseler}},
  \bibinfo {author} {\bibfnamefont {C.}~\bibnamefont {Moritz}}, \bibinfo
  {author} {\bibfnamefont {C.}~\bibnamefont {Dellago}}, \bibinfo {author}
  {\bibfnamefont {R.}~\bibnamefont {Quidant}}, \ and\ \bibinfo {author}
  {\bibfnamefont {L.}~\bibnamefont {Novotny}},\ }\href {\doibase
  10.1103/PhysRevLett.116.243601} {\bibfield  {journal} {\bibinfo  {journal}
  {Phys. Rev. Lett.}\ }\textbf {\bibinfo {volume} {116}},\ \bibinfo {pages}
  {243601} (\bibinfo {year} {2016})}\BibitemShut {NoStop}%
\bibitem [{\citenamefont {Kaltenbaek}\ \emph {et~al.}(2012)\citenamefont
  {Kaltenbaek}, \citenamefont {Hechenblaikner}, \citenamefont {Kiesel},
  \citenamefont {Romero-Isart}, \citenamefont {Schwab}, \citenamefont
  {Johann},\ and\ \citenamefont {Aspelmeyer}}]{Kaltenbaek2012}%
  \BibitemOpen
  \bibfield  {author} {\bibinfo {author} {\bibfnamefont {R.}~\bibnamefont
  {Kaltenbaek}}, \bibinfo {author} {\bibfnamefont {G.}~\bibnamefont
  {Hechenblaikner}}, \bibinfo {author} {\bibfnamefont {N.}~\bibnamefont
  {Kiesel}}, \bibinfo {author} {\bibfnamefont {O.}~\bibnamefont
  {Romero-Isart}}, \bibinfo {author} {\bibfnamefont {K.~C.}\ \bibnamefont
  {Schwab}}, \bibinfo {author} {\bibfnamefont {U.}~\bibnamefont {Johann}}, \
  and\ \bibinfo {author} {\bibfnamefont {M.}~\bibnamefont {Aspelmeyer}},\
  }\href {\doibase 10.1007/s10686-012-9292-3} {\bibfield  {journal} {\bibinfo
  {journal} {Experimental Astronomy}\ }\textbf {\bibinfo {volume} {34}},\
  \bibinfo {pages} {123} (\bibinfo {year} {2012})}\BibitemShut {NoStop}%
\bibitem [{\citenamefont {Romero-Isart}(2011)}]{Romero-Isart2011b}%
  \BibitemOpen
  \bibfield  {author} {\bibinfo {author} {\bibfnamefont {O.}~\bibnamefont
  {Romero-Isart}},\ }\href {\doibase 10.1103/PhysRevA.84.052121} {\bibfield
  {journal} {\bibinfo  {journal} {Phys. Rev. A}\ }\textbf {\bibinfo {volume}
  {84}},\ \bibinfo {pages} {052121} (\bibinfo {year} {2011})}\BibitemShut
  {NoStop}%
\bibitem [{\citenamefont {Kiesel}\ \emph {et~al.}(2013)\citenamefont {Kiesel},
  \citenamefont {Blaser}, \citenamefont {Deli\'{c}}, \citenamefont {Grass},
  \citenamefont {Kaltenbaek},\ and\ \citenamefont {Aspelmeyer}}]{Kiesel2013}%
  \BibitemOpen
  \bibfield  {author} {\bibinfo {author} {\bibfnamefont {N.}~\bibnamefont
  {Kiesel}}, \bibinfo {author} {\bibfnamefont {F.}~\bibnamefont {Blaser}},
  \bibinfo {author} {\bibfnamefont {U.}~\bibnamefont {Deli\'{c}}}, \bibinfo
  {author} {\bibfnamefont {D.}~\bibnamefont {Grass}}, \bibinfo {author}
  {\bibfnamefont {R.}~\bibnamefont {Kaltenbaek}}, \ and\ \bibinfo {author}
  {\bibfnamefont {M.}~\bibnamefont {Aspelmeyer}},\ }\href {\doibase
  10.1073/pnas.1309167110} {\bibfield  {journal} {\bibinfo  {journal} {Proc.
  Natl. Acad. Sci. USA}\ }\textbf {\bibinfo {volume} {110}},\ \bibinfo {pages}
  {14180} (\bibinfo {year} {2013})}\BibitemShut {NoStop}%
\bibitem [{\citenamefont {Windey}\ \emph {et~al.}(2019)\citenamefont {Windey},
  \citenamefont {Gonzalez-Ballestero}, \citenamefont {Maurer}, \citenamefont
  {Novotny}, \citenamefont {Romero-Isart},\ and\ \citenamefont
  {Reimann}}]{Windey2019}%
  \BibitemOpen
  \bibfield  {author} {\bibinfo {author} {\bibfnamefont {D.}~\bibnamefont
  {Windey}}, \bibinfo {author} {\bibfnamefont {C.}~\bibnamefont
  {Gonzalez-Ballestero}}, \bibinfo {author} {\bibfnamefont {P.}~\bibnamefont
  {Maurer}}, \bibinfo {author} {\bibfnamefont {L.}~\bibnamefont {Novotny}},
  \bibinfo {author} {\bibfnamefont {O.}~\bibnamefont {Romero-Isart}}, \ and\
  \bibinfo {author} {\bibfnamefont {R.}~\bibnamefont {Reimann}},\ }\href
  {\doibase 10.1103/PhysRevLett.122.123601} {\bibfield  {journal} {\bibinfo
  {journal} {Phys. Rev. Lett.}\ }\textbf {\bibinfo {volume} {122}},\ \bibinfo
  {pages} {123601} (\bibinfo {year} {2019})}\BibitemShut {NoStop}%
\bibitem [{\citenamefont {Deli{\'c}}\ \emph {et~al.}(2019)\citenamefont
  {Deli{\'c}}, \citenamefont {Reisenbauer}, \citenamefont {Grass},
  \citenamefont {Kiesel}, \citenamefont {Vuleti{\'c}},\ and\ \citenamefont
  {Aspelmeyer}}]{Delic2019}%
  \BibitemOpen
  \bibfield  {author} {\bibinfo {author} {\bibfnamefont {U.}~\bibnamefont
  {Deli{\'c}}}, \bibinfo {author} {\bibfnamefont {M.}~\bibnamefont
  {Reisenbauer}}, \bibinfo {author} {\bibfnamefont {D.}~\bibnamefont {Grass}},
  \bibinfo {author} {\bibfnamefont {N.}~\bibnamefont {Kiesel}}, \bibinfo
  {author} {\bibfnamefont {V.}~\bibnamefont {Vuleti{\'c}}}, \ and\ \bibinfo
  {author} {\bibfnamefont {M.}~\bibnamefont {Aspelmeyer}},\ }\href {\doibase
  10.1103/PhysRevLett.122.123602} {\bibfield  {journal} {\bibinfo  {journal}
  {Phys. Rev. Lett.}\ }\textbf {\bibinfo {volume} {122}},\ \bibinfo {pages}
  {123602} (\bibinfo {year} {2019})}\BibitemShut {NoStop}%
\bibitem [{\citenamefont {Deli{\'c}}\ \emph {et~al.}(2020)\citenamefont
  {Deli{\'c}}, \citenamefont {Reisenbauer}, \citenamefont {Dare}, \citenamefont
  {Grass}, \citenamefont {Vuleti{\'c}}, \citenamefont {Kiesel},\ and\
  \citenamefont {Aspelmeyer}}]{Delic2020}%
  \BibitemOpen
  \bibfield  {author} {\bibinfo {author} {\bibfnamefont {U.}~\bibnamefont
  {Deli{\'c}}}, \bibinfo {author} {\bibfnamefont {M.}~\bibnamefont
  {Reisenbauer}}, \bibinfo {author} {\bibfnamefont {K.}~\bibnamefont {Dare}},
  \bibinfo {author} {\bibfnamefont {D.}~\bibnamefont {Grass}}, \bibinfo
  {author} {\bibfnamefont {V.}~\bibnamefont {Vuleti{\'c}}}, \bibinfo {author}
  {\bibfnamefont {N.}~\bibnamefont {Kiesel}}, \ and\ \bibinfo {author}
  {\bibfnamefont {M.}~\bibnamefont {Aspelmeyer}},\ }\href {\doibase
  10.1126/science.aba3993} {\bibfield  {journal} {\bibinfo  {journal}
  {Science}\ }\textbf {\bibinfo {volume} {367}},\ \bibinfo {pages} {892}
  (\bibinfo {year} {2020})}\BibitemShut {NoStop}%
\bibitem [{\citenamefont {Li}\ \emph {et~al.}(2011)\citenamefont {Li},
  \citenamefont {Kheifets},\ and\ \citenamefont {Raizen}}]{Li2011}%
  \BibitemOpen
  \bibfield  {author} {\bibinfo {author} {\bibfnamefont {T.}~\bibnamefont
  {Li}}, \bibinfo {author} {\bibfnamefont {S.}~\bibnamefont {Kheifets}}, \ and\
  \bibinfo {author} {\bibfnamefont {M.~G.}\ \bibnamefont {Raizen}},\
  }\href@noop {} {\bibfield  {journal} {\bibinfo  {journal} {Nat. Phys.}\
  }\textbf {\bibinfo {volume} {7}},\ \bibinfo {pages} {527} (\bibinfo {year}
  {2011})}\BibitemShut {NoStop}%
\bibitem [{\citenamefont {Gieseler}\ \emph {et~al.}(2012)\citenamefont
  {Gieseler}, \citenamefont {Deutsch}, \citenamefont {Quidant},\ and\
  \citenamefont {Novotny}}]{Gieseler2012}%
  \BibitemOpen
  \bibfield  {author} {\bibinfo {author} {\bibfnamefont {J.}~\bibnamefont
  {Gieseler}}, \bibinfo {author} {\bibfnamefont {B.}~\bibnamefont {Deutsch}},
  \bibinfo {author} {\bibfnamefont {R.}~\bibnamefont {Quidant}}, \ and\
  \bibinfo {author} {\bibfnamefont {L.}~\bibnamefont {Novotny}},\ }\href
  {\doibase 10.1103/PhysRevLett.109.103603} {\bibfield  {journal} {\bibinfo
  {journal} {Phys. Rev. Lett.}\ }\textbf {\bibinfo {volume} {109}},\ \bibinfo
  {pages} {103603} (\bibinfo {year} {2012})}\BibitemShut {NoStop}%
\bibitem [{\citenamefont {Iwasaki}\ \emph {et~al.}(2019)\citenamefont
  {Iwasaki}, \citenamefont {Yotsuya}, \citenamefont {Naruki}, \citenamefont
  {Matsuda}, \citenamefont {Yoneda},\ and\ \citenamefont
  {Aikawa}}]{Iwasaki2019}%
  \BibitemOpen
  \bibfield  {author} {\bibinfo {author} {\bibfnamefont {M.}~\bibnamefont
  {Iwasaki}}, \bibinfo {author} {\bibfnamefont {T.}~\bibnamefont {Yotsuya}},
  \bibinfo {author} {\bibfnamefont {T.}~\bibnamefont {Naruki}}, \bibinfo
  {author} {\bibfnamefont {Y.}~\bibnamefont {Matsuda}}, \bibinfo {author}
  {\bibfnamefont {M.}~\bibnamefont {Yoneda}}, \ and\ \bibinfo {author}
  {\bibfnamefont {K.}~\bibnamefont {Aikawa}},\ }\href {\doibase
  10.1103/PhysRevA.99.051401} {\bibfield  {journal} {\bibinfo  {journal} {Phys.
  Rev. A}\ }\textbf {\bibinfo {volume} {99}},\ \bibinfo {pages} {051401}
  (\bibinfo {year} {2019})}\BibitemShut {NoStop}%
\bibitem [{\citenamefont {Conangla}\ \emph {et~al.}(2019)\citenamefont
  {Conangla}, \citenamefont {Ricci}, \citenamefont {Cuairan}, \citenamefont
  {Schell}, \citenamefont {Meyer},\ and\ \citenamefont
  {Quidant}}]{Conangela2019}%
  \BibitemOpen
  \bibfield  {author} {\bibinfo {author} {\bibfnamefont {G.~P.}\ \bibnamefont
  {Conangla}}, \bibinfo {author} {\bibfnamefont {F.}~\bibnamefont {Ricci}},
  \bibinfo {author} {\bibfnamefont {M.~T.}\ \bibnamefont {Cuairan}}, \bibinfo
  {author} {\bibfnamefont {A.~W.}\ \bibnamefont {Schell}}, \bibinfo {author}
  {\bibfnamefont {N.}~\bibnamefont {Meyer}}, \ and\ \bibinfo {author}
  {\bibfnamefont {R.}~\bibnamefont {Quidant}},\ }\href {\doibase
  10.1103/PhysRevLett.122.223602} {\bibfield  {journal} {\bibinfo  {journal}
  {Phys. Rev. Lett.}\ }\textbf {\bibinfo {volume} {122}},\ \bibinfo {pages}
  {223602} (\bibinfo {year} {2019})}\BibitemShut {NoStop}%
\bibitem [{\citenamefont {Tebbenjohanns}\ \emph
  {et~al.}(2019{\natexlab{a}})\citenamefont {Tebbenjohanns}, \citenamefont
  {Frimmer}, \citenamefont {Militaru}, \citenamefont {Jain},\ and\
  \citenamefont {Novotny}}]{Tebbenjohanns2019}%
  \BibitemOpen
  \bibfield  {author} {\bibinfo {author} {\bibfnamefont {F.}~\bibnamefont
  {Tebbenjohanns}}, \bibinfo {author} {\bibfnamefont {M.}~\bibnamefont
  {Frimmer}}, \bibinfo {author} {\bibfnamefont {A.}~\bibnamefont {Militaru}},
  \bibinfo {author} {\bibfnamefont {V.}~\bibnamefont {Jain}}, \ and\ \bibinfo
  {author} {\bibfnamefont {L.}~\bibnamefont {Novotny}},\ }\href {\doibase
  10.1103/PhysRevLett.122.223601} {\bibfield  {journal} {\bibinfo  {journal}
  {Phys. Rev. Lett.}\ }\textbf {\bibinfo {volume} {122}},\ \bibinfo {pages}
  {223601} (\bibinfo {year} {2019}{\natexlab{a}})}\BibitemShut {NoStop}%
\bibitem [{\citenamefont {Mancini}\ \emph {et~al.}(1998)\citenamefont
  {Mancini}, \citenamefont {Vitali},\ and\ \citenamefont
  {Tombesi}}]{Mancini1998}%
  \BibitemOpen
  \bibfield  {author} {\bibinfo {author} {\bibfnamefont {S.}~\bibnamefont
  {Mancini}}, \bibinfo {author} {\bibfnamefont {D.}~\bibnamefont {Vitali}}, \
  and\ \bibinfo {author} {\bibfnamefont {P.}~\bibnamefont {Tombesi}},\ }\href
  {\doibase 10.1103/PhysRevLett.80.688} {\bibfield  {journal} {\bibinfo
  {journal} {Phys. Rev. Lett.}\ }\textbf {\bibinfo {volume} {80}},\ \bibinfo
  {pages} {688} (\bibinfo {year} {1998})}\BibitemShut {NoStop}%
\bibitem [{\citenamefont {Wilson}\ \emph {et~al.}(2015)\citenamefont {Wilson},
  \citenamefont {Sudhir}, \citenamefont {Piro}, \citenamefont {Schilling},
  \citenamefont {Ghadimi},\ and\ \citenamefont {Kippenberg}}]{Wilson2015}%
  \BibitemOpen
  \bibfield  {author} {\bibinfo {author} {\bibfnamefont {D.}~\bibnamefont
  {Wilson}}, \bibinfo {author} {\bibfnamefont {V.}~\bibnamefont {Sudhir}},
  \bibinfo {author} {\bibfnamefont {N.}~\bibnamefont {Piro}}, \bibinfo {author}
  {\bibfnamefont {R.}~\bibnamefont {Schilling}}, \bibinfo {author}
  {\bibfnamefont {A.}~\bibnamefont {Ghadimi}}, \ and\ \bibinfo {author}
  {\bibfnamefont {T.~J.}\ \bibnamefont {Kippenberg}},\ }\href
  {http://www.nature.com/nature/journal/v524/n7565/full/nature14672.html}
  {\bibfield  {journal} {\bibinfo  {journal} {Nature}\ }\textbf {\bibinfo
  {volume} {524}},\ \bibinfo {pages} {325} (\bibinfo {year}
  {2015})}\BibitemShut {NoStop}%
\bibitem [{\citenamefont {Cohadon}\ \emph {et~al.}(1999)\citenamefont
  {Cohadon}, \citenamefont {Heidmann},\ and\ \citenamefont
  {Pinard}}]{Cohadon1999}%
  \BibitemOpen
  \bibfield  {author} {\bibinfo {author} {\bibfnamefont {P.~F.}\ \bibnamefont
  {Cohadon}}, \bibinfo {author} {\bibfnamefont {A.}~\bibnamefont {Heidmann}}, \
  and\ \bibinfo {author} {\bibfnamefont {M.}~\bibnamefont {Pinard}},\ }\href
  {\doibase 10.1103/PhysRevLett.83.3174} {\bibfield  {journal} {\bibinfo
  {journal} {Phys. Rev. Lett.}\ }\textbf {\bibinfo {volume} {83}},\ \bibinfo
  {pages} {3174} (\bibinfo {year} {1999})}\BibitemShut {NoStop}%
\bibitem [{\citenamefont {Poggio}\ \emph {et~al.}(2007)\citenamefont {Poggio},
  \citenamefont {Degen}, \citenamefont {Mamin},\ and\ \citenamefont
  {Rugar}}]{Poggio2007}%
  \BibitemOpen
  \bibfield  {author} {\bibinfo {author} {\bibfnamefont {M.}~\bibnamefont
  {Poggio}}, \bibinfo {author} {\bibfnamefont {C.~L.}\ \bibnamefont {Degen}},
  \bibinfo {author} {\bibfnamefont {H.~J.}\ \bibnamefont {Mamin}}, \ and\
  \bibinfo {author} {\bibfnamefont {D.}~\bibnamefont {Rugar}},\ }\href
  {\doibase 10.1103/PhysRevLett.99.017201} {\bibfield  {journal} {\bibinfo
  {journal} {Phys. Rev. Lett.}\ }\textbf {\bibinfo {volume} {99}},\ \bibinfo
  {pages} {017201} (\bibinfo {year} {2007})}\BibitemShut {NoStop}%
\bibitem [{\citenamefont {Kamba}\ \emph {et~al.}(2020)\citenamefont {Kamba},
  \citenamefont {Kiuchi}, \citenamefont {Yotsuya},\ and\ \citenamefont
  {Aikawa}}]{Kamba2020}%
  \BibitemOpen
  \bibfield  {author} {\bibinfo {author} {\bibfnamefont {M.}~\bibnamefont
  {Kamba}}, \bibinfo {author} {\bibfnamefont {H.}~\bibnamefont {Kiuchi}},
  \bibinfo {author} {\bibfnamefont {T.}~\bibnamefont {Yotsuya}}, \ and\
  \bibinfo {author} {\bibfnamefont {K.}~\bibnamefont {Aikawa}},\ }\href@noop {}
  {\  (\bibinfo {year} {2020})},\ \Eprint
  {http://arxiv.org/abs/arXiv:2011.12507} {arXiv:2011.12507} \BibitemShut
  {NoStop}%
\bibitem [{\citenamefont {Tebbenjohanns}\ \emph {et~al.}(2020)\citenamefont
  {Tebbenjohanns}, \citenamefont {Frimmer}, \citenamefont {Jain}, \citenamefont
  {Windey},\ and\ \citenamefont {Novotny}}]{Tebbenjohanns2020}%
  \BibitemOpen
  \bibfield  {author} {\bibinfo {author} {\bibfnamefont {F.}~\bibnamefont
  {Tebbenjohanns}}, \bibinfo {author} {\bibfnamefont {M.}~\bibnamefont
  {Frimmer}}, \bibinfo {author} {\bibfnamefont {V.}~\bibnamefont {Jain}},
  \bibinfo {author} {\bibfnamefont {D.}~\bibnamefont {Windey}}, \ and\ \bibinfo
  {author} {\bibfnamefont {L.}~\bibnamefont {Novotny}},\ }\href {\doibase
  10.1103/PhysRevLett.124.013603} {\bibfield  {journal} {\bibinfo  {journal}
  {Phys. Rev. Lett.}\ }\textbf {\bibinfo {volume} {124}},\ \bibinfo {pages}
  {013603} (\bibinfo {year} {2020})}\BibitemShut {NoStop}%
\bibitem [{\citenamefont {Tebbenjohanns}\ \emph
  {et~al.}(2019{\natexlab{b}})\citenamefont {Tebbenjohanns}, \citenamefont
  {Frimmer},\ and\ \citenamefont {Novotny}}]{Tebbenjohanns2019Efficiency}%
  \BibitemOpen
  \bibfield  {author} {\bibinfo {author} {\bibfnamefont {F.}~\bibnamefont
  {Tebbenjohanns}}, \bibinfo {author} {\bibfnamefont {M.}~\bibnamefont
  {Frimmer}}, \ and\ \bibinfo {author} {\bibfnamefont {L.}~\bibnamefont
  {Novotny}},\ }\href {\doibase 10.1103/PhysRevA.100.043821} {\bibfield
  {journal} {\bibinfo  {journal} {Phys. Rev. A}\ }\textbf {\bibinfo {volume}
  {100}},\ \bibinfo {pages} {043821} (\bibinfo {year}
  {2019}{\natexlab{b}})}\BibitemShut {NoStop}%
\bibitem [{\citenamefont {Genes}\ \emph {et~al.}(2008)\citenamefont {Genes},
  \citenamefont {Vitali}, \citenamefont {Tombesi}, \citenamefont {Gigan},\ and\
  \citenamefont {Aspelmeyer}}]{Genes2008}%
  \BibitemOpen
  \bibfield  {author} {\bibinfo {author} {\bibfnamefont {C.}~\bibnamefont
  {Genes}}, \bibinfo {author} {\bibfnamefont {D.}~\bibnamefont {Vitali}},
  \bibinfo {author} {\bibfnamefont {P.}~\bibnamefont {Tombesi}}, \bibinfo
  {author} {\bibfnamefont {S.}~\bibnamefont {Gigan}}, \ and\ \bibinfo {author}
  {\bibfnamefont {M.}~\bibnamefont {Aspelmeyer}},\ }\href {\doibase
  10.1103/PhysRevA.77.033804} {\bibfield  {journal} {\bibinfo  {journal} {Phys.
  Rev. A}\ }\textbf {\bibinfo {volume} {77}},\ \bibinfo {pages} {033804}
  (\bibinfo {year} {2008})}\BibitemShut {NoStop}%
\bibitem [{\citenamefont {Clerk}\ \emph {et~al.}(2010)\citenamefont {Clerk},
  \citenamefont {Devoret}, \citenamefont {Girvin}, \citenamefont {Marquardt},\
  and\ \citenamefont {Schoelkopf}}]{Clerk2010}%
  \BibitemOpen
  \bibfield  {author} {\bibinfo {author} {\bibfnamefont {A.~A.}\ \bibnamefont
  {Clerk}}, \bibinfo {author} {\bibfnamefont {M.~H.}\ \bibnamefont {Devoret}},
  \bibinfo {author} {\bibfnamefont {S.~M.}\ \bibnamefont {Girvin}}, \bibinfo
  {author} {\bibfnamefont {F.}~\bibnamefont {Marquardt}}, \ and\ \bibinfo
  {author} {\bibfnamefont {R.~J.}\ \bibnamefont {Schoelkopf}},\ }\href
  {\doibase 10.1103/RevModPhys.82.1155} {\bibfield  {journal} {\bibinfo
  {journal} {Rev. Mod. Phys.}\ }\textbf {\bibinfo {volume} {82}},\ \bibinfo
  {pages} {1155} (\bibinfo {year} {2010})}\BibitemShut {NoStop}%
\bibitem [{\citenamefont {Safavi-Naeini}\ \emph {et~al.}(2012)\citenamefont
  {Safavi-Naeini}, \citenamefont {Chan}, \citenamefont {Hill}, \citenamefont
  {Alegre}, \citenamefont {Krause},\ and\ \citenamefont
  {Painter}}]{Safavi-Naeini2012}%
  \BibitemOpen
  \bibfield  {author} {\bibinfo {author} {\bibfnamefont {A.~H.}\ \bibnamefont
  {Safavi-Naeini}}, \bibinfo {author} {\bibfnamefont {J.}~\bibnamefont {Chan}},
  \bibinfo {author} {\bibfnamefont {J.~T.}\ \bibnamefont {Hill}}, \bibinfo
  {author} {\bibfnamefont {T.~P.~M.}\ \bibnamefont {Alegre}}, \bibinfo {author}
  {\bibfnamefont {A.}~\bibnamefont {Krause}}, \ and\ \bibinfo {author}
  {\bibfnamefont {O.}~\bibnamefont {Painter}},\ }\href {\doibase
  10.1103/PhysRevLett.108.033602} {\bibfield  {journal} {\bibinfo  {journal}
  {Phys. Rev. Lett.}\ }\textbf {\bibinfo {volume} {108}},\ \bibinfo {pages}
  {033602} (\bibinfo {year} {2012})}\BibitemShut {NoStop}%
\bibitem [{Note1()}]{Note1}%
  \BibitemOpen
  \bibinfo {note} {We define our two-sided, symmetrized PSDs $\bar
  {S}_{zz}(\protect \ensuremath {\Omega })$ and our single-sided PSDs $\protect
  \tilde {S}_{zz}(f)=4\pi \bar {S}_{zz}(2\pi f)$ according to $\langle z^2
  \rangle = \DOTSI \intop \ilimits@ _{-\infty }^\infty \protect \text
  {d}\protect \ensuremath {\Omega }~ \bar {S}_{zz}(\protect \ensuremath {\Omega
  })= \DOTSI \intop \ilimits@ _0^\infty \protect \text {d}f~\protect \tilde
  {S}_{zz}(f)$}\BibitemShut {NoStop}%
\bibitem [{\citenamefont {Purdy}\ \emph {et~al.}(2017)\citenamefont {Purdy},
  \citenamefont {Grutter}, \citenamefont {Srinivasan},\ and\ \citenamefont
  {Taylor}}]{Purdy2017}%
  \BibitemOpen
  \bibfield  {author} {\bibinfo {author} {\bibfnamefont {T.~P.}\ \bibnamefont
  {Purdy}}, \bibinfo {author} {\bibfnamefont {K.~E.}\ \bibnamefont {Grutter}},
  \bibinfo {author} {\bibfnamefont {K.}~\bibnamefont {Srinivasan}}, \ and\
  \bibinfo {author} {\bibfnamefont {J.~M.}\ \bibnamefont {Taylor}},\ }\href
  {\doibase 10.1126/science.aag1407} {\bibfield  {journal} {\bibinfo  {journal}
  {Science}\ }\textbf {\bibinfo {volume} {356}},\ \bibinfo {pages} {1265}
  (\bibinfo {year} {2017})}\BibitemShut {NoStop}%
\bibitem [{\citenamefont {Shkarin}\ \emph {et~al.}(2019)\citenamefont
  {Shkarin}, \citenamefont {Kashkanova}, \citenamefont {Brown}, \citenamefont
  {Garcia}, \citenamefont {Ott}, \citenamefont {Reichel},\ and\ \citenamefont
  {Harris}}]{Shkarin2019}%
  \BibitemOpen
  \bibfield  {author} {\bibinfo {author} {\bibfnamefont {A.~B.}\ \bibnamefont
  {Shkarin}}, \bibinfo {author} {\bibfnamefont {A.~D.}\ \bibnamefont
  {Kashkanova}}, \bibinfo {author} {\bibfnamefont {C.~D.}\ \bibnamefont
  {Brown}}, \bibinfo {author} {\bibfnamefont {S.}~\bibnamefont {Garcia}},
  \bibinfo {author} {\bibfnamefont {K.}~\bibnamefont {Ott}}, \bibinfo {author}
  {\bibfnamefont {J.}~\bibnamefont {Reichel}}, \ and\ \bibinfo {author}
  {\bibfnamefont {J.~G.~E.}\ \bibnamefont {Harris}},\ }\href {\doibase
  10.1103/PhysRevLett.122.153601} {\bibfield  {journal} {\bibinfo  {journal}
  {Phys. Rev. Lett.}\ }\textbf {\bibinfo {volume} {122}},\ \bibinfo {pages}
  {153601} (\bibinfo {year} {2019})}\BibitemShut {NoStop}%
\bibitem [{\citenamefont {Wiseman}\ and\ \citenamefont
  {Milburn}(2010)}]{Wiseman2010}%
  \BibitemOpen
  \bibfield  {author} {\bibinfo {author} {\bibfnamefont {H.~M.}\ \bibnamefont
  {Wiseman}}\ and\ \bibinfo {author} {\bibfnamefont {G.~J.}\ \bibnamefont
  {Milburn}},\ }\href@noop {} {\emph {\bibinfo {title} {Quantum Measurement and
  Control}}}\ (\bibinfo  {publisher} {{Cambridge University Press}},\ \bibinfo
  {year} {2010})\BibitemShut {NoStop}%
\bibitem [{\citenamefont {Braginsky}\ and\ \citenamefont
  {Khalili}(1992)}]{Braginsky1992}%
  \BibitemOpen
  \bibfield  {author} {\bibinfo {author} {\bibfnamefont {V.~B.}\ \bibnamefont
  {Braginsky}}\ and\ \bibinfo {author} {\bibfnamefont {F.~Y.}\ \bibnamefont
  {Khalili}},\ }\href@noop {} {\emph {\bibinfo {title} {Quantum Measurement}}}\
  (\bibinfo  {publisher} {Cambridge University Press, Cambridge},\ \bibinfo
  {year} {1992})\BibitemShut {NoStop}%
\bibitem [{\citenamefont {Sayrin}\ \emph {et~al.}(2011)\citenamefont {Sayrin},
  \citenamefont {Dotsenko}, \citenamefont {Zhou}, \citenamefont {Peaudecerf},
  \citenamefont {Rybarczyk}, \citenamefont {Gleyzes}, \citenamefont {Rouchon},
  \citenamefont {Mirrahimi}, \citenamefont {Amini}, \citenamefont {Brune},\
  and\ \citenamefont {et~al.}}]{Sayrin2011}%
  \BibitemOpen
  \bibfield  {author} {\bibinfo {author} {\bibfnamefont {C.}~\bibnamefont
  {Sayrin}}, \bibinfo {author} {\bibfnamefont {I.}~\bibnamefont {Dotsenko}},
  \bibinfo {author} {\bibfnamefont {X.}~\bibnamefont {Zhou}}, \bibinfo {author}
  {\bibfnamefont {B.}~\bibnamefont {Peaudecerf}}, \bibinfo {author}
  {\bibfnamefont {T.}~\bibnamefont {Rybarczyk}}, \bibinfo {author}
  {\bibfnamefont {S.}~\bibnamefont {Gleyzes}}, \bibinfo {author} {\bibfnamefont
  {P.}~\bibnamefont {Rouchon}}, \bibinfo {author} {\bibfnamefont
  {M.}~\bibnamefont {Mirrahimi}}, \bibinfo {author} {\bibfnamefont
  {H.}~\bibnamefont {Amini}}, \bibinfo {author} {\bibfnamefont
  {M.}~\bibnamefont {Brune}}, \ and\ \bibinfo {author} {\bibnamefont
  {et~al.}},\ }\href {\doibase 10.1038/nature10376} {\bibfield  {journal}
  {\bibinfo  {journal} {Nature}\ }\textbf {\bibinfo {volume} {477}},\ \bibinfo
  {pages} {73–77} (\bibinfo {year} {2011})}\BibitemShut {NoStop}%
\bibitem [{\citenamefont {Vijay}\ \emph {et~al.}(2012)\citenamefont {Vijay},
  \citenamefont {Macklin}, \citenamefont {Slichter}, \citenamefont {Weber},
  \citenamefont {Murch}, \citenamefont {Naik}, \citenamefont {Korotkov},\ and\
  \citenamefont {Siddiqi}}]{Vijay2012}%
  \BibitemOpen
  \bibfield  {author} {\bibinfo {author} {\bibfnamefont {R.}~\bibnamefont
  {Vijay}}, \bibinfo {author} {\bibfnamefont {C.}~\bibnamefont {Macklin}},
  \bibinfo {author} {\bibfnamefont {D.~H.}\ \bibnamefont {Slichter}}, \bibinfo
  {author} {\bibfnamefont {S.~J.}\ \bibnamefont {Weber}}, \bibinfo {author}
  {\bibfnamefont {K.~W.}\ \bibnamefont {Murch}}, \bibinfo {author}
  {\bibfnamefont {R.}~\bibnamefont {Naik}}, \bibinfo {author} {\bibfnamefont
  {A.~N.}\ \bibnamefont {Korotkov}}, \ and\ \bibinfo {author} {\bibfnamefont
  {I.}~\bibnamefont {Siddiqi}},\ }\href {\doibase 10.1038/nature11505}
  {\bibfield  {journal} {\bibinfo  {journal} {Nature}\ }\textbf {\bibinfo
  {volume} {490}},\ \bibinfo {pages} {77–80} (\bibinfo {year}
  {2012})}\BibitemShut {NoStop}%
\bibitem [{\citenamefont {Doherty}\ and\ \citenamefont
  {Jacobs}(1999)}]{Doherty1999b}%
  \BibitemOpen
  \bibfield  {author} {\bibinfo {author} {\bibfnamefont {A.~C.}\ \bibnamefont
  {Doherty}}\ and\ \bibinfo {author} {\bibfnamefont {K.}~\bibnamefont
  {Jacobs}},\ }\href {\doibase 10.1103/PhysRevA.60.2700} {\bibfield  {journal}
  {\bibinfo  {journal} {Phys. Rev. A}\ }\textbf {\bibinfo {volume} {60}},\
  \bibinfo {pages} {2700} (\bibinfo {year} {1999})}\BibitemShut {NoStop}%
\bibitem [{\citenamefont {Magrini}\ \emph {et~al.}(2020)\citenamefont
  {Magrini}, \citenamefont {Rosenzweig}, \citenamefont {Bach}, \citenamefont
  {Deutschmann-Olek}, \citenamefont {Hofer}, \citenamefont {Hong},
  \citenamefont {Kiesel}, \citenamefont {Kugi},\ and\ \citenamefont
  {Aspelmeyer}}]{Magrini2020}%
  \BibitemOpen
  \bibfield  {author} {\bibinfo {author} {\bibfnamefont {L.}~\bibnamefont
  {Magrini}}, \bibinfo {author} {\bibfnamefont {P.}~\bibnamefont {Rosenzweig}},
  \bibinfo {author} {\bibfnamefont {C.}~\bibnamefont {Bach}}, \bibinfo {author}
  {\bibfnamefont {A.}~\bibnamefont {Deutschmann-Olek}}, \bibinfo {author}
  {\bibfnamefont {S.~G.}\ \bibnamefont {Hofer}}, \bibinfo {author}
  {\bibfnamefont {S.}~\bibnamefont {Hong}}, \bibinfo {author} {\bibfnamefont
  {N.}~\bibnamefont {Kiesel}}, \bibinfo {author} {\bibfnamefont
  {A.}~\bibnamefont {Kugi}}, \ and\ \bibinfo {author} {\bibfnamefont
  {M.}~\bibnamefont {Aspelmeyer}},\ }\href@noop {} {\enquote {\bibinfo {title}
  {Optimal quantum control of mechanical motion at room temperature:
  ground-state cooling},}\ } (\bibinfo {year} {2020}),\ \Eprint
  {http://arxiv.org/abs/2012.15188} {arXiv:2012.15188 [quant-ph]} \BibitemShut
  {NoStop}%
\bibitem [{\citenamefont {Meng}\ \emph {et~al.}(2020)\citenamefont {Meng},
  \citenamefont {Brawley}, \citenamefont {Bennett}, \citenamefont {Vanner},\
  and\ \citenamefont {Bowen}}]{Meng2020}%
  \BibitemOpen
  \bibfield  {author} {\bibinfo {author} {\bibfnamefont {C.}~\bibnamefont
  {Meng}}, \bibinfo {author} {\bibfnamefont {G.~A.}\ \bibnamefont {Brawley}},
  \bibinfo {author} {\bibfnamefont {J.~S.}\ \bibnamefont {Bennett}}, \bibinfo
  {author} {\bibfnamefont {M.~R.}\ \bibnamefont {Vanner}}, \ and\ \bibinfo
  {author} {\bibfnamefont {W.~P.}\ \bibnamefont {Bowen}},\ }\href {\doibase
  10.1103/PhysRevLett.125.043604} {\bibfield  {journal} {\bibinfo  {journal}
  {Phys. Rev. Lett.}\ }\textbf {\bibinfo {volume} {125}},\ \bibinfo {pages}
  {043604} (\bibinfo {year} {2020})}\BibitemShut {NoStop}%
\bibitem [{\citenamefont {Vanner}\ \emph {et~al.}(2011)\citenamefont {Vanner},
  \citenamefont {Pikovski}, \citenamefont {Cole}, \citenamefont {Kim},
  \citenamefont {Brukner}, \citenamefont {Hammerer}, \citenamefont {Milburn},\
  and\ \citenamefont {Aspelmeyer}}]{Vanner2011}%
  \BibitemOpen
  \bibfield  {author} {\bibinfo {author} {\bibfnamefont {M.~R.}\ \bibnamefont
  {Vanner}}, \bibinfo {author} {\bibfnamefont {I.}~\bibnamefont {Pikovski}},
  \bibinfo {author} {\bibfnamefont {G.~D.}\ \bibnamefont {Cole}}, \bibinfo
  {author} {\bibfnamefont {M.~S.}\ \bibnamefont {Kim}}, \bibinfo {author}
  {\bibfnamefont {{\v C}.}~\bibnamefont {Brukner}}, \bibinfo {author}
  {\bibfnamefont {K.}~\bibnamefont {Hammerer}}, \bibinfo {author}
  {\bibfnamefont {G.~J.}\ \bibnamefont {Milburn}}, \ and\ \bibinfo {author}
  {\bibfnamefont {M.}~\bibnamefont {Aspelmeyer}},\ }\href {\doibase
  10.1073/pnas.1105098108} {\bibfield  {journal} {\bibinfo  {journal} {Proc.
  Natl. Acad. Sci. USA}\ }\textbf {\bibinfo {volume} {108}},\ \bibinfo {pages}
  {16182} (\bibinfo {year} {2011})}\BibitemShut {NoStop}%
\bibitem [{\citenamefont {Gabrielse}\ \emph {et~al.}(1990)\citenamefont
  {Gabrielse}, \citenamefont {Fei}, \citenamefont {Orozco}, \citenamefont
  {Tjoelker}, \citenamefont {Haas}, \citenamefont {Kalinowsky}, \citenamefont
  {Trainor},\ and\ \citenamefont {Kells}}]{Gabrielse1990}%
  \BibitemOpen
  \bibfield  {author} {\bibinfo {author} {\bibfnamefont {G.}~\bibnamefont
  {Gabrielse}}, \bibinfo {author} {\bibfnamefont {X.}~\bibnamefont {Fei}},
  \bibinfo {author} {\bibfnamefont {L.~A.}\ \bibnamefont {Orozco}}, \bibinfo
  {author} {\bibfnamefont {R.~L.}\ \bibnamefont {Tjoelker}}, \bibinfo {author}
  {\bibfnamefont {J.}~\bibnamefont {Haas}}, \bibinfo {author} {\bibfnamefont
  {H.}~\bibnamefont {Kalinowsky}}, \bibinfo {author} {\bibfnamefont {T.~A.}\
  \bibnamefont {Trainor}}, \ and\ \bibinfo {author} {\bibfnamefont
  {W.}~\bibnamefont {Kells}},\ }\href {\doibase 10.1103/PhysRevLett.65.1317}
  {\bibfield  {journal} {\bibinfo  {journal} {Phys. Rev. Lett.}\ }\textbf
  {\bibinfo {volume} {65}},\ \bibinfo {pages} {1317} (\bibinfo {year}
  {1990})}\BibitemShut {NoStop}%
\bibitem [{\citenamefont {Bateman}\ \emph {et~al.}(2014)\citenamefont
  {Bateman}, \citenamefont {Nimmrichter}, \citenamefont {Hornberger},\ and\
  \citenamefont {Ulbricht}}]{Bateman2014}%
  \BibitemOpen
  \bibfield  {author} {\bibinfo {author} {\bibfnamefont {J.}~\bibnamefont
  {Bateman}}, \bibinfo {author} {\bibfnamefont {S.}~\bibnamefont
  {Nimmrichter}}, \bibinfo {author} {\bibfnamefont {K.}~\bibnamefont
  {Hornberger}}, \ and\ \bibinfo {author} {\bibfnamefont {H.}~\bibnamefont
  {Ulbricht}},\ }\href {\doibase 10.1038/ncomms5788} {\bibfield  {journal}
  {\bibinfo  {journal} {Nat. Commun.}\ }\textbf {\bibinfo {volume} {5}},\
  \bibinfo {pages} {4788} (\bibinfo {year} {2014})}\BibitemShut {NoStop}%
\bibitem [{\citenamefont {Doherty}\ \emph {et~al.}(1999)\citenamefont
  {Doherty}, \citenamefont {Tan}, \citenamefont {Parkins},\ and\ \citenamefont
  {Walls}}]{Doherty1999}%
  \BibitemOpen
  \bibfield  {author} {\bibinfo {author} {\bibfnamefont {A.~C.}\ \bibnamefont
  {Doherty}}, \bibinfo {author} {\bibfnamefont {S.~M.}\ \bibnamefont {Tan}},
  \bibinfo {author} {\bibfnamefont {A.~S.}\ \bibnamefont {Parkins}}, \ and\
  \bibinfo {author} {\bibfnamefont {D.~F.}\ \bibnamefont {Walls}},\ }\href
  {\doibase 10.1103/PhysRevA.60.2380} {\bibfield  {journal} {\bibinfo
  {journal} {Phys. Rev. A}\ }\textbf {\bibinfo {volume} {60}},\ \bibinfo
  {pages} {2380} (\bibinfo {year} {1999})}\BibitemShut {NoStop}%
\bibitem [{\citenamefont {Micke}\ \emph {et~al.}(2019)\citenamefont {Micke},
  \citenamefont {Stark}, \citenamefont {King}, \citenamefont {Leopold},
  \citenamefont {Pfeifer}, \citenamefont {Schmöger}, \citenamefont {Schwarz},
  \citenamefont {Spieß}, \citenamefont {Schmidt}, \citenamefont
  {López-Urrutia},\ and\ \citenamefont {et~al.}}]{Micke2019}%
  \BibitemOpen
  \bibfield  {author} {\bibinfo {author} {\bibfnamefont {P.}~\bibnamefont
  {Micke}}, \bibinfo {author} {\bibfnamefont {J.}~\bibnamefont {Stark}},
  \bibinfo {author} {\bibfnamefont {S.~A.}\ \bibnamefont {King}}, \bibinfo
  {author} {\bibfnamefont {T.}~\bibnamefont {Leopold}}, \bibinfo {author}
  {\bibfnamefont {T.}~\bibnamefont {Pfeifer}}, \bibinfo {author} {\bibfnamefont
  {L.}~\bibnamefont {Schmöger}}, \bibinfo {author} {\bibfnamefont
  {M.}~\bibnamefont {Schwarz}}, \bibinfo {author} {\bibfnamefont {L.~J.}\
  \bibnamefont {Spieß}}, \bibinfo {author} {\bibfnamefont {P.~O.}\
  \bibnamefont {Schmidt}}, \bibinfo {author} {\bibfnamefont {J.~R.~C.}\
  \bibnamefont {López-Urrutia}}, \ and\ \bibinfo {author} {\bibnamefont
  {et~al.}},\ }\href {\doibase 10.1063/1.5088593} {\bibfield  {journal}
  {\bibinfo  {journal} {Rev. Sci. Instr.}\ }\textbf {\bibinfo {volume} {90}},\
  \bibinfo {pages} {065104} (\bibinfo {year} {2019})}\BibitemShut {NoStop}%
\bibitem [{\citenamefont {Kaltenbaek}\ \emph {et~al.}(2016)\citenamefont
  {Kaltenbaek}, \citenamefont {Aspelmeyer}, \citenamefont {Barker},
  \citenamefont {Bassi}, \citenamefont {Bateman}, \citenamefont {Bongs},
  \citenamefont {Bose}, \citenamefont {Braxmaier}, \citenamefont {Brukner},
  \citenamefont {Christophe}, \citenamefont {Chwalla}, \citenamefont {Cohadon},
  \citenamefont {Cruise}, \citenamefont {Curceanu}, \citenamefont {Dholakia},
  \citenamefont {Di{\'o}si}, \citenamefont {D{\"o}ringshoff}, \citenamefont
  {Ertmer}, \citenamefont {Gieseler}, \citenamefont {G{\"u}rlebeck},
  \citenamefont {Hechenblaikner}, \citenamefont {Heidmann}, \citenamefont
  {Herrmann}, \citenamefont {Hossenfelder}, \citenamefont {Johann},
  \citenamefont {Kiesel}, \citenamefont {Kim}, \citenamefont {L{\"a}mmerzahl},
  \citenamefont {Lambrecht}, \citenamefont {Mazilu}, \citenamefont {Milburn},
  \citenamefont {M{\"u}ller}, \citenamefont {Novotny}, \citenamefont
  {Paternostro}, \citenamefont {Peters}, \citenamefont {Pikovski},
  \citenamefont {Pilan~Zanoni}, \citenamefont {Rasel}, \citenamefont {Reynaud},
  \citenamefont {Riedel}, \citenamefont {Rodrigues}, \citenamefont {Rondin},
  \citenamefont {Roura}, \citenamefont {Schleich}, \citenamefont
  {Schmiedmayer}, \citenamefont {Schuldt}, \citenamefont {Schwab},
  \citenamefont {Tajmar}, \citenamefont {Tino}, \citenamefont {Ulbricht},
  \citenamefont {Ursin},\ and\ \citenamefont {Vedral}}]{Kaltenbaek2016}%
  \BibitemOpen
  \bibfield  {author} {\bibinfo {author} {\bibfnamefont {R.}~\bibnamefont
  {Kaltenbaek}}, \bibinfo {author} {\bibfnamefont {M.}~\bibnamefont
  {Aspelmeyer}}, \bibinfo {author} {\bibfnamefont {P.~F.}\ \bibnamefont
  {Barker}}, \bibinfo {author} {\bibfnamefont {A.}~\bibnamefont {Bassi}},
  \bibinfo {author} {\bibfnamefont {J.}~\bibnamefont {Bateman}}, \bibinfo
  {author} {\bibfnamefont {K.}~\bibnamefont {Bongs}}, \bibinfo {author}
  {\bibfnamefont {S.}~\bibnamefont {Bose}}, \bibinfo {author} {\bibfnamefont
  {C.}~\bibnamefont {Braxmaier}}, \bibinfo {author} {\bibfnamefont {{\v
  C}.}~\bibnamefont {Brukner}}, \bibinfo {author} {\bibfnamefont
  {B.}~\bibnamefont {Christophe}}, \bibinfo {author} {\bibfnamefont
  {M.}~\bibnamefont {Chwalla}}, \bibinfo {author} {\bibfnamefont {P.-F.}\
  \bibnamefont {Cohadon}}, \bibinfo {author} {\bibfnamefont {A.~M.}\
  \bibnamefont {Cruise}}, \bibinfo {author} {\bibfnamefont {C.}~\bibnamefont
  {Curceanu}}, \bibinfo {author} {\bibfnamefont {K.}~\bibnamefont {Dholakia}},
  \bibinfo {author} {\bibfnamefont {L.}~\bibnamefont {Di{\'o}si}}, \bibinfo
  {author} {\bibfnamefont {K.}~\bibnamefont {D{\"o}ringshoff}}, \bibinfo
  {author} {\bibfnamefont {W.}~\bibnamefont {Ertmer}}, \bibinfo {author}
  {\bibfnamefont {J.}~\bibnamefont {Gieseler}}, \bibinfo {author}
  {\bibfnamefont {N.}~\bibnamefont {G{\"u}rlebeck}}, \bibinfo {author}
  {\bibfnamefont {G.}~\bibnamefont {Hechenblaikner}}, \bibinfo {author}
  {\bibfnamefont {A.}~\bibnamefont {Heidmann}}, \bibinfo {author}
  {\bibfnamefont {S.}~\bibnamefont {Herrmann}}, \bibinfo {author}
  {\bibfnamefont {S.}~\bibnamefont {Hossenfelder}}, \bibinfo {author}
  {\bibfnamefont {U.}~\bibnamefont {Johann}}, \bibinfo {author} {\bibfnamefont
  {N.}~\bibnamefont {Kiesel}}, \bibinfo {author} {\bibfnamefont
  {M.}~\bibnamefont {Kim}}, \bibinfo {author} {\bibfnamefont {C.}~\bibnamefont
  {L{\"a}mmerzahl}}, \bibinfo {author} {\bibfnamefont {A.}~\bibnamefont
  {Lambrecht}}, \bibinfo {author} {\bibfnamefont {M.}~\bibnamefont {Mazilu}},
  \bibinfo {author} {\bibfnamefont {G.~J.}\ \bibnamefont {Milburn}}, \bibinfo
  {author} {\bibfnamefont {H.}~\bibnamefont {M{\"u}ller}}, \bibinfo {author}
  {\bibfnamefont {L.}~\bibnamefont {Novotny}}, \bibinfo {author} {\bibfnamefont
  {M.}~\bibnamefont {Paternostro}}, \bibinfo {author} {\bibfnamefont
  {A.}~\bibnamefont {Peters}}, \bibinfo {author} {\bibfnamefont
  {I.}~\bibnamefont {Pikovski}}, \bibinfo {author} {\bibfnamefont
  {A.}~\bibnamefont {Pilan~Zanoni}}, \bibinfo {author} {\bibfnamefont {E.~M.}\
  \bibnamefont {Rasel}}, \bibinfo {author} {\bibfnamefont {S.}~\bibnamefont
  {Reynaud}}, \bibinfo {author} {\bibfnamefont {C.~J.}\ \bibnamefont {Riedel}},
  \bibinfo {author} {\bibfnamefont {M.}~\bibnamefont {Rodrigues}}, \bibinfo
  {author} {\bibfnamefont {L.}~\bibnamefont {Rondin}}, \bibinfo {author}
  {\bibfnamefont {A.}~\bibnamefont {Roura}}, \bibinfo {author} {\bibfnamefont
  {W.~P.}\ \bibnamefont {Schleich}}, \bibinfo {author} {\bibfnamefont
  {J.}~\bibnamefont {Schmiedmayer}}, \bibinfo {author} {\bibfnamefont
  {T.}~\bibnamefont {Schuldt}}, \bibinfo {author} {\bibfnamefont {K.~C.}\
  \bibnamefont {Schwab}}, \bibinfo {author} {\bibfnamefont {M.}~\bibnamefont
  {Tajmar}}, \bibinfo {author} {\bibfnamefont {G.~M.}\ \bibnamefont {Tino}},
  \bibinfo {author} {\bibfnamefont {H.}~\bibnamefont {Ulbricht}}, \bibinfo
  {author} {\bibfnamefont {R.}~\bibnamefont {Ursin}}, \ and\ \bibinfo {author}
  {\bibfnamefont {V.}~\bibnamefont {Vedral}},\ }\href {\doibase
  10.1140/epjqt/s40507-016-0043-7} {\bibfield  {journal} {\bibinfo  {journal}
  {EPJ Quantum Technol.}\ }\textbf {\bibinfo {volume} {3}},\ \bibinfo {pages}
  {5} (\bibinfo {year} {2016})}\BibitemShut {NoStop}%
\end{thebibliography}%


\begin{thebibliography}{22}%
\makeatletter
\providecommand \@ifxundefined [1]{%
 \@ifx{#1\undefined}
}%
\providecommand \@ifnum [1]{%
 \ifnum #1\expandafter \@firstoftwo
 \else \expandafter \@secondoftwo
 \fi
}%
\providecommand \@ifx [1]{%
 \ifx #1\expandafter \@firstoftwo
 \else \expandafter \@secondoftwo
 \fi
}%
\providecommand \natexlab [1]{#1}%
\providecommand \enquote  [1]{``#1''}%
\providecommand \bibnamefont  [1]{#1}%
\providecommand \bibfnamefont [1]{#1}%
\providecommand \citenamefont [1]{#1}%
\providecommand \href@noop [0]{\@secondoftwo}%
\providecommand \href [0]{\begingroup \@sanitize@url \@href}%
\providecommand \@href[1]{\@@startlink{#1}\@@href}%
\providecommand \@@href[1]{\endgroup#1\@@endlink}%
\providecommand \@sanitize@url [0]{\catcode `\\12\catcode `\$12\catcode
  `\&12\catcode `\#12\catcode `\^12\catcode `\_12\catcode `\%12\relax}%
\providecommand \@@startlink[1]{}%
\providecommand \@@endlink[0]{}%
\providecommand \url  [0]{\begingroup\@sanitize@url \@url }%
\providecommand \@url [1]{\endgroup\@href {#1}{\urlprefix }}%
\providecommand \urlprefix  [0]{URL }%
\providecommand \Eprint [0]{\href }%
\providecommand \doibase [0]{http://dx.doi.org/}%
\providecommand \selectlanguage [0]{\@gobble}%
\providecommand \bibinfo  [0]{\@secondoftwo}%
\providecommand \bibfield  [0]{\@secondoftwo}%
\providecommand \translation [1]{[#1]}%
\providecommand \BibitemOpen [0]{}%
\providecommand \bibitemStop [0]{}%
\providecommand \bibitemNoStop [0]{.\EOS\space}%
\providecommand \EOS [0]{\spacefactor3000\relax}%
\providecommand \BibitemShut  [1]{\csname bibitem#1\endcsname}%
\let\auto@bib@innerbib\@empty
\bibitem [{\citenamefont {Micke}\ \emph {et~al.}(2019)\citenamefont {Micke},
  \citenamefont {Stark}, \citenamefont {King}, \citenamefont {Leopold},
  \citenamefont {Pfeifer}, \citenamefont {Schmöger}, \citenamefont {Schwarz},
  \citenamefont {Spieß}, \citenamefont {Schmidt}, \citenamefont
  {López-Urrutia},\ and\ \citenamefont {et~al.}}]{Micke2019}%
  \BibitemOpen
  \bibfield  {author} {\bibinfo {author} {\bibfnamefont {P.}~\bibnamefont
  {Micke}}, \bibinfo {author} {\bibfnamefont {J.}~\bibnamefont {Stark}},
  \bibinfo {author} {\bibfnamefont {S.~A.}\ \bibnamefont {King}}, \bibinfo
  {author} {\bibfnamefont {T.}~\bibnamefont {Leopold}}, \bibinfo {author}
  {\bibfnamefont {T.}~\bibnamefont {Pfeifer}}, \bibinfo {author} {\bibfnamefont
  {L.}~\bibnamefont {Schmöger}}, \bibinfo {author} {\bibfnamefont
  {M.}~\bibnamefont {Schwarz}}, \bibinfo {author} {\bibfnamefont {L.~J.}\
  \bibnamefont {Spieß}}, \bibinfo {author} {\bibfnamefont {P.~O.}\
  \bibnamefont {Schmidt}}, \bibinfo {author} {\bibfnamefont {J.~R.~C.}\
  \bibnamefont {López-Urrutia}}, \ and\ \bibinfo {author} {\bibnamefont
  {et~al.}},\ }\href {\doibase 10.1063/1.5088593} {\bibfield  {journal}
  {\bibinfo  {journal} {Rev. Sci. Instr.}\ }\textbf {\bibinfo {volume} {90}},\
  \bibinfo {pages} {065104} (\bibinfo {year} {2019})}\BibitemShut {NoStop}%
\bibitem [{\citenamefont {Ahn}\ \emph {et~al.}(2018)\citenamefont {Ahn},
  \citenamefont {Xu}, \citenamefont {Bang}, \citenamefont {Deng}, \citenamefont
  {Hoang}, \citenamefont {Han}, \citenamefont {Ma},\ and\ \citenamefont
  {Li}}]{Ahn2018}%
  \BibitemOpen
  \bibfield  {author} {\bibinfo {author} {\bibfnamefont {J.}~\bibnamefont
  {Ahn}}, \bibinfo {author} {\bibfnamefont {Z.}~\bibnamefont {Xu}}, \bibinfo
  {author} {\bibfnamefont {J.}~\bibnamefont {Bang}}, \bibinfo {author}
  {\bibfnamefont {Y.-H.}\ \bibnamefont {Deng}}, \bibinfo {author}
  {\bibfnamefont {T.~M.}\ \bibnamefont {Hoang}}, \bibinfo {author}
  {\bibfnamefont {Q.}~\bibnamefont {Han}}, \bibinfo {author} {\bibfnamefont
  {R.-M.}\ \bibnamefont {Ma}}, \ and\ \bibinfo {author} {\bibfnamefont
  {T.}~\bibnamefont {Li}},\ }\href {\doibase 10.1103/PhysRevLett.121.033603}
  {\bibfield  {journal} {\bibinfo  {journal} {Phys. Rev. Lett.}\ }\textbf
  {\bibinfo {volume} {121}},\ \bibinfo {pages} {033603} (\bibinfo {year}
  {2018})}\BibitemShut {NoStop}%
\bibitem [{\citenamefont {Hebestreit}\ \emph {et~al.}(2018)\citenamefont
  {Hebestreit}, \citenamefont {Frimmer}, \citenamefont {Reimann}, \citenamefont
  {Dellago}, \citenamefont {Ricci},\ and\ \citenamefont
  {Novotny}}]{Hebestreit2018}%
  \BibitemOpen
  \bibfield  {author} {\bibinfo {author} {\bibfnamefont {E.}~\bibnamefont
  {Hebestreit}}, \bibinfo {author} {\bibfnamefont {M.}~\bibnamefont {Frimmer}},
  \bibinfo {author} {\bibfnamefont {R.}~\bibnamefont {Reimann}}, \bibinfo
  {author} {\bibfnamefont {C.}~\bibnamefont {Dellago}}, \bibinfo {author}
  {\bibfnamefont {F.}~\bibnamefont {Ricci}}, \ and\ \bibinfo {author}
  {\bibfnamefont {L.}~\bibnamefont {Novotny}},\ }\href {\doibase
  10.1063/1.5017119} {\bibfield  {journal} {\bibinfo  {journal} {Rev. Sci.
  Instr.}\ }\textbf {\bibinfo {volume} {89}},\ \bibinfo {pages} {033111}
  (\bibinfo {year} {2018})}\BibitemShut {NoStop}%
\bibitem [{\citenamefont {van~der Laan}\ \emph {et~al.}(2020)\citenamefont
  {van~der Laan}, \citenamefont {Reimann}, \citenamefont {Militaru},
  \citenamefont {Tebbenjohanns}, \citenamefont {Windey}, \citenamefont
  {Frimmer},\ and\ \citenamefont {Novotny}}]{Laan2020}%
  \BibitemOpen
  \bibfield  {author} {\bibinfo {author} {\bibfnamefont {F.}~\bibnamefont
  {van~der Laan}}, \bibinfo {author} {\bibfnamefont {R.}~\bibnamefont
  {Reimann}}, \bibinfo {author} {\bibfnamefont {A.}~\bibnamefont {Militaru}},
  \bibinfo {author} {\bibfnamefont {F.}~\bibnamefont {Tebbenjohanns}}, \bibinfo
  {author} {\bibfnamefont {D.}~\bibnamefont {Windey}}, \bibinfo {author}
  {\bibfnamefont {M.}~\bibnamefont {Frimmer}}, \ and\ \bibinfo {author}
  {\bibfnamefont {L.}~\bibnamefont {Novotny}},\ }\href {\doibase
  10.1103/PhysRevA.102.013505} {\bibfield  {journal} {\bibinfo  {journal}
  {Phys. Rev. A}\ }\textbf {\bibinfo {volume} {102}},\ \bibinfo {pages}
  {013505} (\bibinfo {year} {2020})}\BibitemShut {NoStop}%
\bibitem [{\citenamefont {Gieseler}\ \emph {et~al.}(2012)\citenamefont
  {Gieseler}, \citenamefont {Deutsch}, \citenamefont {Quidant},\ and\
  \citenamefont {Novotny}}]{Gieseler2012}%
  \BibitemOpen
  \bibfield  {author} {\bibinfo {author} {\bibfnamefont {J.}~\bibnamefont
  {Gieseler}}, \bibinfo {author} {\bibfnamefont {B.}~\bibnamefont {Deutsch}},
  \bibinfo {author} {\bibfnamefont {R.}~\bibnamefont {Quidant}}, \ and\
  \bibinfo {author} {\bibfnamefont {L.}~\bibnamefont {Novotny}},\ }\href
  {\doibase 10.1103/PhysRevLett.109.103603} {\bibfield  {journal} {\bibinfo
  {journal} {Phys. Rev. Lett.}\ }\textbf {\bibinfo {volume} {109}},\ \bibinfo
  {pages} {103603} (\bibinfo {year} {2012})}\BibitemShut {NoStop}%
\bibitem [{\citenamefont {Jain}\ \emph {et~al.}(2016)\citenamefont {Jain},
  \citenamefont {Gieseler}, \citenamefont {Moritz}, \citenamefont {Dellago},
  \citenamefont {Quidant},\ and\ \citenamefont {Novotny}}]{Jain2016}%
  \BibitemOpen
  \bibfield  {author} {\bibinfo {author} {\bibfnamefont {V.}~\bibnamefont
  {Jain}}, \bibinfo {author} {\bibfnamefont {J.}~\bibnamefont {Gieseler}},
  \bibinfo {author} {\bibfnamefont {C.}~\bibnamefont {Moritz}}, \bibinfo
  {author} {\bibfnamefont {C.}~\bibnamefont {Dellago}}, \bibinfo {author}
  {\bibfnamefont {R.}~\bibnamefont {Quidant}}, \ and\ \bibinfo {author}
  {\bibfnamefont {L.}~\bibnamefont {Novotny}},\ }\href {\doibase
  10.1103/PhysRevLett.116.243601} {\bibfield  {journal} {\bibinfo  {journal}
  {Phys. Rev. Lett.}\ }\textbf {\bibinfo {volume} {116}},\ \bibinfo {pages}
  {243601} (\bibinfo {year} {2016})}\BibitemShut {NoStop}%
\bibitem [{\citenamefont {Underwood}\ \emph {et~al.}(2015)\citenamefont
  {Underwood}, \citenamefont {Mason}, \citenamefont {Lee}, \citenamefont {Xu},
  \citenamefont {Jiang}, \citenamefont {Shkarin}, \citenamefont {B\o{}rkje},
  \citenamefont {Girvin},\ and\ \citenamefont {Harris}}]{Underwood2015}%
  \BibitemOpen
  \bibfield  {author} {\bibinfo {author} {\bibfnamefont {M.}~\bibnamefont
  {Underwood}}, \bibinfo {author} {\bibfnamefont {D.}~\bibnamefont {Mason}},
  \bibinfo {author} {\bibfnamefont {D.}~\bibnamefont {Lee}}, \bibinfo {author}
  {\bibfnamefont {H.}~\bibnamefont {Xu}}, \bibinfo {author} {\bibfnamefont
  {L.}~\bibnamefont {Jiang}}, \bibinfo {author} {\bibfnamefont {A.~B.}\
  \bibnamefont {Shkarin}}, \bibinfo {author} {\bibfnamefont {K.}~\bibnamefont
  {B\o{}rkje}}, \bibinfo {author} {\bibfnamefont {S.~M.}\ \bibnamefont
  {Girvin}}, \ and\ \bibinfo {author} {\bibfnamefont {J.~G.~E.}\ \bibnamefont
  {Harris}},\ }\href {\doibase 10.1103/PhysRevA.92.061801} {\bibfield
  {journal} {\bibinfo  {journal} {Phys. Rev. A}\ }\textbf {\bibinfo {volume}
  {92}},\ \bibinfo {pages} {061801(R)} (\bibinfo {year} {2015})}\BibitemShut
  {NoStop}%
\bibitem [{\citenamefont {Tebbenjohanns}\ \emph {et~al.}(2020)\citenamefont
  {Tebbenjohanns}, \citenamefont {Frimmer}, \citenamefont {Jain}, \citenamefont
  {Windey},\ and\ \citenamefont {Novotny}}]{Tebbenjohanns2020}%
  \BibitemOpen
  \bibfield  {author} {\bibinfo {author} {\bibfnamefont {F.}~\bibnamefont
  {Tebbenjohanns}}, \bibinfo {author} {\bibfnamefont {M.}~\bibnamefont
  {Frimmer}}, \bibinfo {author} {\bibfnamefont {V.}~\bibnamefont {Jain}},
  \bibinfo {author} {\bibfnamefont {D.}~\bibnamefont {Windey}}, \ and\ \bibinfo
  {author} {\bibfnamefont {L.}~\bibnamefont {Novotny}},\ }\href {\doibase
  10.1103/PhysRevLett.124.013603} {\bibfield  {journal} {\bibinfo  {journal}
  {Phys. Rev. Lett.}\ }\textbf {\bibinfo {volume} {124}},\ \bibinfo {pages}
  {013603} (\bibinfo {year} {2020})}\BibitemShut {NoStop}%
\bibitem [{\citenamefont {Purdy}\ \emph {et~al.}(2017)\citenamefont {Purdy},
  \citenamefont {Grutter}, \citenamefont {Srinivasan},\ and\ \citenamefont
  {Taylor}}]{Purdy2017}%
  \BibitemOpen
  \bibfield  {author} {\bibinfo {author} {\bibfnamefont {T.~P.}\ \bibnamefont
  {Purdy}}, \bibinfo {author} {\bibfnamefont {K.~E.}\ \bibnamefont {Grutter}},
  \bibinfo {author} {\bibfnamefont {K.}~\bibnamefont {Srinivasan}}, \ and\
  \bibinfo {author} {\bibfnamefont {J.~M.}\ \bibnamefont {Taylor}},\ }\href
  {\doibase 10.1126/science.aag1407} {\bibfield  {journal} {\bibinfo  {journal}
  {Science}\ }\textbf {\bibinfo {volume} {356}},\ \bibinfo {pages} {1265}
  (\bibinfo {year} {2017})}\BibitemShut {NoStop}%
\bibitem [{\citenamefont {Shkarin}\ \emph {et~al.}(2019)\citenamefont
  {Shkarin}, \citenamefont {Kashkanova}, \citenamefont {Brown}, \citenamefont
  {Garcia}, \citenamefont {Ott}, \citenamefont {Reichel},\ and\ \citenamefont
  {Harris}}]{Shkarin2019}%
  \BibitemOpen
  \bibfield  {author} {\bibinfo {author} {\bibfnamefont {A.~B.}\ \bibnamefont
  {Shkarin}}, \bibinfo {author} {\bibfnamefont {A.~D.}\ \bibnamefont
  {Kashkanova}}, \bibinfo {author} {\bibfnamefont {C.~D.}\ \bibnamefont
  {Brown}}, \bibinfo {author} {\bibfnamefont {S.}~\bibnamefont {Garcia}},
  \bibinfo {author} {\bibfnamefont {K.}~\bibnamefont {Ott}}, \bibinfo {author}
  {\bibfnamefont {J.}~\bibnamefont {Reichel}}, \ and\ \bibinfo {author}
  {\bibfnamefont {J.~G.~E.}\ \bibnamefont {Harris}},\ }\href {\doibase
  10.1103/PhysRevLett.122.153601} {\bibfield  {journal} {\bibinfo  {journal}
  {Phys. Rev. Lett.}\ }\textbf {\bibinfo {volume} {122}},\ \bibinfo {pages}
  {153601} (\bibinfo {year} {2019})}\BibitemShut {NoStop}%
\bibitem [{\citenamefont {Mancini}\ \emph {et~al.}(1998)\citenamefont
  {Mancini}, \citenamefont {Vitali},\ and\ \citenamefont
  {Tombesi}}]{Mancini1998}%
  \BibitemOpen
  \bibfield  {author} {\bibinfo {author} {\bibfnamefont {S.}~\bibnamefont
  {Mancini}}, \bibinfo {author} {\bibfnamefont {D.}~\bibnamefont {Vitali}}, \
  and\ \bibinfo {author} {\bibfnamefont {P.}~\bibnamefont {Tombesi}},\ }\href
  {\doibase 10.1103/PhysRevLett.80.688} {\bibfield  {journal} {\bibinfo
  {journal} {Phys. Rev. Lett.}\ }\textbf {\bibinfo {volume} {80}},\ \bibinfo
  {pages} {688} (\bibinfo {year} {1998})}\BibitemShut {NoStop}%
\bibitem [{\citenamefont {Doherty}\ and\ \citenamefont
  {Jacobs}(1999)}]{Doherty1999b}%
  \BibitemOpen
  \bibfield  {author} {\bibinfo {author} {\bibfnamefont {A.~C.}\ \bibnamefont
  {Doherty}}\ and\ \bibinfo {author} {\bibfnamefont {K.}~\bibnamefont
  {Jacobs}},\ }\href {\doibase 10.1103/PhysRevA.60.2700} {\bibfield  {journal}
  {\bibinfo  {journal} {Phys. Rev. A}\ }\textbf {\bibinfo {volume} {60}},\
  \bibinfo {pages} {2700} (\bibinfo {year} {1999})}\BibitemShut {NoStop}%
\bibitem [{\citenamefont {Genes}\ \emph {et~al.}(2008)\citenamefont {Genes},
  \citenamefont {Vitali}, \citenamefont {Tombesi}, \citenamefont {Gigan},\ and\
  \citenamefont {Aspelmeyer}}]{Genes2008}%
  \BibitemOpen
  \bibfield  {author} {\bibinfo {author} {\bibfnamefont {C.}~\bibnamefont
  {Genes}}, \bibinfo {author} {\bibfnamefont {D.}~\bibnamefont {Vitali}},
  \bibinfo {author} {\bibfnamefont {P.}~\bibnamefont {Tombesi}}, \bibinfo
  {author} {\bibfnamefont {S.}~\bibnamefont {Gigan}}, \ and\ \bibinfo {author}
  {\bibfnamefont {M.}~\bibnamefont {Aspelmeyer}},\ }\href {\doibase
  10.1103/PhysRevA.77.033804} {\bibfield  {journal} {\bibinfo  {journal} {Phys.
  Rev. A}\ }\textbf {\bibinfo {volume} {77}},\ \bibinfo {pages} {033804}
  (\bibinfo {year} {2008})}\BibitemShut {NoStop}%
\bibitem [{\citenamefont {Cohadon}\ \emph {et~al.}(1999)\citenamefont
  {Cohadon}, \citenamefont {Heidmann},\ and\ \citenamefont
  {Pinard}}]{Cohadon1999}%
  \BibitemOpen
  \bibfield  {author} {\bibinfo {author} {\bibfnamefont {P.~F.}\ \bibnamefont
  {Cohadon}}, \bibinfo {author} {\bibfnamefont {A.}~\bibnamefont {Heidmann}}, \
  and\ \bibinfo {author} {\bibfnamefont {M.}~\bibnamefont {Pinard}},\ }\href
  {\doibase 10.1103/PhysRevLett.83.3174} {\bibfield  {journal} {\bibinfo
  {journal} {Phys. Rev. Lett.}\ }\textbf {\bibinfo {volume} {83}},\ \bibinfo
  {pages} {3174} (\bibinfo {year} {1999})}\BibitemShut {NoStop}%
\bibitem [{\citenamefont {Poggio}\ \emph {et~al.}(2007)\citenamefont {Poggio},
  \citenamefont {Degen}, \citenamefont {Mamin},\ and\ \citenamefont
  {Rugar}}]{Poggio2007}%
  \BibitemOpen
  \bibfield  {author} {\bibinfo {author} {\bibfnamefont {M.}~\bibnamefont
  {Poggio}}, \bibinfo {author} {\bibfnamefont {C.~L.}\ \bibnamefont {Degen}},
  \bibinfo {author} {\bibfnamefont {H.~J.}\ \bibnamefont {Mamin}}, \ and\
  \bibinfo {author} {\bibfnamefont {D.}~\bibnamefont {Rugar}},\ }\href
  {\doibase 10.1103/PhysRevLett.99.017201} {\bibfield  {journal} {\bibinfo
  {journal} {Phys. Rev. Lett.}\ }\textbf {\bibinfo {volume} {99}},\ \bibinfo
  {pages} {017201} (\bibinfo {year} {2007})}\BibitemShut {NoStop}%
\bibitem [{\citenamefont {Wilson}\ \emph {et~al.}(2015)\citenamefont {Wilson},
  \citenamefont {Sudhir}, \citenamefont {Piro}, \citenamefont {Schilling},
  \citenamefont {Ghadimi},\ and\ \citenamefont {Kippenberg}}]{Wilson2015}%
  \BibitemOpen
  \bibfield  {author} {\bibinfo {author} {\bibfnamefont {D.}~\bibnamefont
  {Wilson}}, \bibinfo {author} {\bibfnamefont {V.}~\bibnamefont {Sudhir}},
  \bibinfo {author} {\bibfnamefont {N.}~\bibnamefont {Piro}}, \bibinfo {author}
  {\bibfnamefont {R.}~\bibnamefont {Schilling}}, \bibinfo {author}
  {\bibfnamefont {A.}~\bibnamefont {Ghadimi}}, \ and\ \bibinfo {author}
  {\bibfnamefont {T.~J.}\ \bibnamefont {Kippenberg}},\ }\href
  {http://www.nature.com/nature/journal/v524/n7565/full/nature14672.html}
  {\bibfield  {journal} {\bibinfo  {journal} {Nature}\ }\textbf {\bibinfo
  {volume} {524}},\ \bibinfo {pages} {325} (\bibinfo {year}
  {2015})}\BibitemShut {NoStop}%
\bibitem [{\citenamefont {Rossi}\ \emph {et~al.}(2018)\citenamefont {Rossi},
  \citenamefont {Mason}, \citenamefont {Chen}, \citenamefont {Tsaturyan},\ and\
  \citenamefont {Schliesser}}]{Rossi2018}%
  \BibitemOpen
  \bibfield  {author} {\bibinfo {author} {\bibfnamefont {M.}~\bibnamefont
  {Rossi}}, \bibinfo {author} {\bibfnamefont {D.}~\bibnamefont {Mason}},
  \bibinfo {author} {\bibfnamefont {J.}~\bibnamefont {Chen}}, \bibinfo {author}
  {\bibfnamefont {Y.}~\bibnamefont {Tsaturyan}}, \ and\ \bibinfo {author}
  {\bibfnamefont {A.}~\bibnamefont {Schliesser}},\ }\href {\doibase
  10.1038/s41586-018-0643-8} {\bibfield  {journal} {\bibinfo  {journal}
  {Nature}\ }\textbf {\bibinfo {volume} {563}},\ \bibinfo {pages} {53–58}
  (\bibinfo {year} {2018})}\BibitemShut {NoStop}%
\bibitem [{\citenamefont {Wiseman}\ and\ \citenamefont
  {Milburn}(2010)}]{Wiseman2010}%
  \BibitemOpen
  \bibfield  {author} {\bibinfo {author} {\bibfnamefont {H.~M.}\ \bibnamefont
  {Wiseman}}\ and\ \bibinfo {author} {\bibfnamefont {G.~J.}\ \bibnamefont
  {Milburn}},\ }\href@noop {} {\emph {\bibinfo {title} {Quantum Measurement and
  Control}}}\ (\bibinfo  {publisher} {{Cambridge University Press}},\ \bibinfo
  {year} {2010})\BibitemShut {NoStop}%
\bibitem [{\citenamefont {Rodenburg}\ \emph {et~al.}(2016)\citenamefont
  {Rodenburg}, \citenamefont {Neukirch}, \citenamefont {Vamivakas},\ and\
  \citenamefont {Bhattacharya}}]{Rodenburg2016}%
  \BibitemOpen
  \bibfield  {author} {\bibinfo {author} {\bibfnamefont {B.}~\bibnamefont
  {Rodenburg}}, \bibinfo {author} {\bibfnamefont {L.~P.}\ \bibnamefont
  {Neukirch}}, \bibinfo {author} {\bibfnamefont {A.~N.}\ \bibnamefont
  {Vamivakas}}, \ and\ \bibinfo {author} {\bibfnamefont {M.}~\bibnamefont
  {Bhattacharya}},\ }\href {\doibase 10.1364/OPTICA.3.000318} {\bibfield
  {journal} {\bibinfo  {journal} {Optica}\ }\textbf {\bibinfo {volume} {3}},\
  \bibinfo {pages} {318} (\bibinfo {year} {2016})}\BibitemShut {NoStop}%
\bibitem [{\citenamefont {Wieczorek}\ \emph {et~al.}(2015)\citenamefont
  {Wieczorek}, \citenamefont {Hofer}, \citenamefont {{Hoelscher-Obermaier}},
  \citenamefont {Riedinger}, \citenamefont {Hammerer},\ and\ \citenamefont
  {Aspelmeyer}}]{Wieczorek2015}%
  \BibitemOpen
  \bibfield  {author} {\bibinfo {author} {\bibfnamefont {W.}~\bibnamefont
  {Wieczorek}}, \bibinfo {author} {\bibfnamefont {S.~G.}\ \bibnamefont
  {Hofer}}, \bibinfo {author} {\bibfnamefont {J.}~\bibnamefont
  {{Hoelscher-Obermaier}}}, \bibinfo {author} {\bibfnamefont {R.}~\bibnamefont
  {Riedinger}}, \bibinfo {author} {\bibfnamefont {K.}~\bibnamefont {Hammerer}},
  \ and\ \bibinfo {author} {\bibfnamefont {M.}~\bibnamefont {Aspelmeyer}},\
  }\href {\doibase 10.1103/PhysRevLett.114.223601} {\bibfield  {journal}
  {\bibinfo  {journal} {Phys. Rev. Lett.}\ }\textbf {\bibinfo {volume} {114}},\
  \bibinfo {pages} {223601} (\bibinfo {year} {2015})}\BibitemShut {NoStop}%
\bibitem [{\citenamefont {Garbini}\ \emph {et~al.}(1996)\citenamefont
  {Garbini}, \citenamefont {Bruland}, \citenamefont {Dougherty},\ and\
  \citenamefont {Sidles}}]{Garbini1996}%
  \BibitemOpen
  \bibfield  {author} {\bibinfo {author} {\bibfnamefont {J.~L.}\ \bibnamefont
  {Garbini}}, \bibinfo {author} {\bibfnamefont {K.~J.}\ \bibnamefont
  {Bruland}}, \bibinfo {author} {\bibfnamefont {W.~M.}\ \bibnamefont
  {Dougherty}}, \ and\ \bibinfo {author} {\bibfnamefont {J.~A.}\ \bibnamefont
  {Sidles}},\ }\href {\doibase 10.1063/1.363085} {\bibfield  {journal}
  {\bibinfo  {journal} {J. Appl. Phys.}\ }\textbf {\bibinfo {volume} {80}},\
  \bibinfo {pages} {1951} (\bibinfo {year} {1996})}\BibitemShut {NoStop}%
\bibitem [{\citenamefont {Aspelmeyer}\ \emph {et~al.}(2014)\citenamefont
  {Aspelmeyer}, \citenamefont {Kippenberg},\ and\ \citenamefont
  {Marquardt}}]{Aspelmeyer2014}%
  \BibitemOpen
  \bibfield  {author} {\bibinfo {author} {\bibfnamefont {M.}~\bibnamefont
  {Aspelmeyer}}, \bibinfo {author} {\bibfnamefont {T.~J.}\ \bibnamefont
  {Kippenberg}}, \ and\ \bibinfo {author} {\bibfnamefont {F.}~\bibnamefont
  {Marquardt}},\ }\href {\doibase 10.1103/RevModPhys.86.1391} {\bibfield
  {journal} {\bibinfo  {journal} {Rev. Mod. Phys.}\ }\textbf {\bibinfo {volume}
  {86}},\ \bibinfo {pages} {1391} (\bibinfo {year} {2014})}\BibitemShut
  {NoStop}%
\end{thebibliography}%

\end{document}


\title{Supplementary Information for\\Quantum control of a nanoparticle optically levitated in cryogenic free space}

\author{Felix Tebbenjohanns}
\altaffiliation{\equalcontribution}
\affiliation{\affil}
\author{M. Luisa Mattana}
\altaffiliation{\equalcontribution}
\affiliation{\affil}
\author{Massimiliano Rossi}
\altaffiliation{\equalcontribution}
\affiliation{\affil}
\author{Martin Frimmer}
\affiliation{\affil}
\author{Lukas Novotny}
\affiliation{\affil}
\homepage{http://www.photonics.ethz.ch}


\maketitle
\tableofcontents

\newpage
\section{Notation}
Throughout this work, we use the following symbols.

\begin{tabular}{c c}
Symbol & Explanation \\ \hline
   $m$  & Mass of the particle\\
    $\Oz$ & Eigenfrequency of the particle along the optical axis \\
    $\zzpf$ & Zero-point fluctuation size: $\zzpf = \sqrt{\hbar/(2m\Oz)}$ \\
    $\Geff$ & Effective feedback-induced mechanical damping rate \\
    $\gamma_\text{th}$ & Damping rate due to thermal bath: $\gamma_\text{th} \ll \Geff$ \\
    $T$ & Bath temperature \\
    $\Gth$ & Thermal decoherence rate (phonons/s): $\Gth = \gamma_\text{th} k_B T/(\hbar \Oz)$ \\
    $\Gqba$ & Decoherence rate due to quantum backaction \\
    $\Gexc$ & Excess decoherence rate including $\Gth$ \\
    $\Gtot$ & Total decoherence rate: $\Gtot = \Gqba+ \Gexc = \Gqba(1+1/C_q)$  \\
    $C_q$ & Quantum cooperativity: $C_q = \Gqba/\Gexc$ \\
    $\etadet$ & Detection efficiency \\
    $\Gmeas$ & Measurement rate: $\Gmeas = \eta_d \Gqba $ \\
    $\etameas$ & Measurement efficiency: $\etameas = \Gmeas/\Gtot = \etadet/(1+1/C_q)$ \\
    $\nbar$ & Phonon occupation number of the particle's z-motion \\
    $\chi_\text{eff}(\Og)$ & Effective mechanical susceptibility: $\chi_\text{eff}(\Og)=m^{-1}/(\Oz^2-\Og^2-\imu\Geff\Og)$  \\
    $\Sfftot$ & Two-sided, symmetrized total force noise PSD: $\Sfftot = \hbar^2\Gtot/(2\pi\zzpf^2) $ \\
    $\Simp$ & Detector imprecision noise PSD: $\Simp = \zzpf^2/(8\pi\Gmeas)$ \\
    $\Simp\Sfftot $ & Measurement-disturbance relation: $\Simp\Sfftot  = (\hbar/4\pi)^2/\etameas$ \\
    $\Sbar_{zz}(\Og)$ & Two-sided, symmetrized particle position PSD: $\Braket{z^2} = \int \diff \Og~\Sbar_{zz}(\Og) $ \\
    $\Sbar_{zz}^\text{hom}(\Og)$ & Measured in-loop position PSD on the homodyne detector \\
    $\Sbar_{rr}(\Og)$ & Heterodyne detector position PSD at the Stokes (red) sideband \\
    $\Sbar_{bb}(\Og)$ & Heterodyne detector position PSD at the anti-Stokes (blue) sideband  \\
    $\Sbar_{rb}(\Og)$ & Cross-PSD between Stokes and anti-Stokes sidebands
\end{tabular}

\section{Experimental methods}
\begin{figure}[tb]
\includegraphics[width=\textwidth]{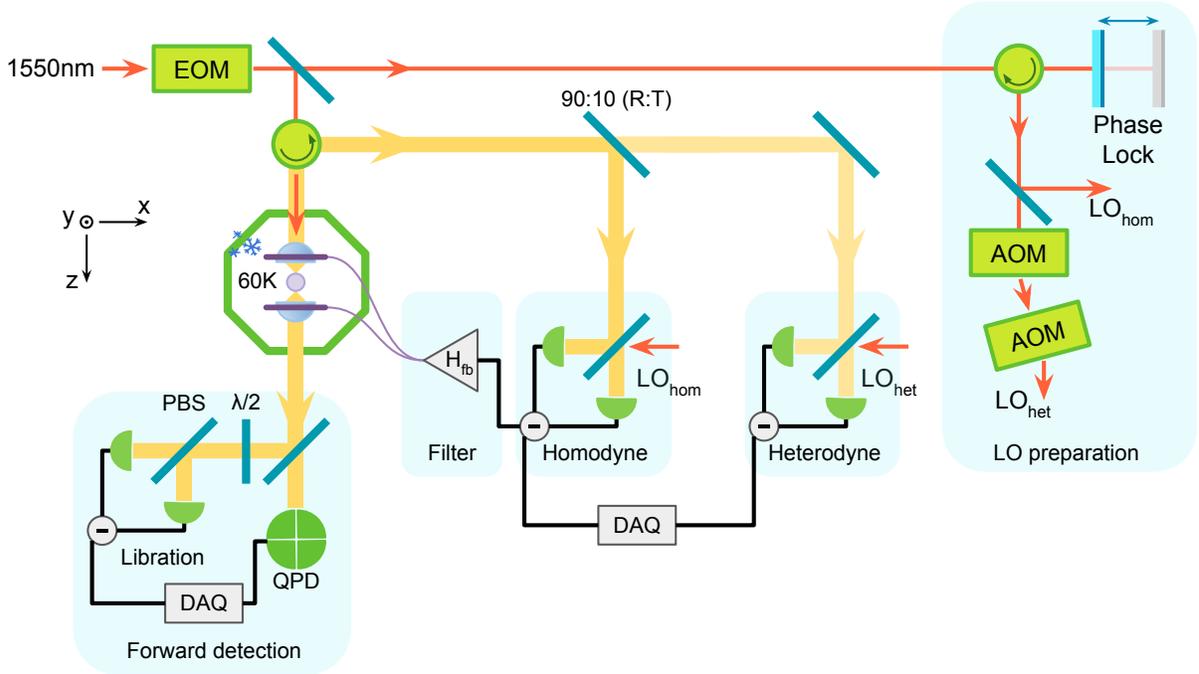}
\caption{{\bf Experimental Setup.} 
We optically trap a nanoparticle inside a cryogenic vacuum chamber using a telecom laser. In the forward direction, we employ a libration and position detection system. In the backward direction, we place both a homodyne and a heterodyne photodetector. 
AOM: acousto-optic modulator. DAQ: data acquisition card. EOM: electro-optic modulator. $\lambda/2$: half-wave plate. LO: local oscillator. PBS: polarizing beam-splitter.  R: reflection. T: transmission.
}
\label{fig:setup_detailed}
\end{figure}

\subsection{Setup}
    
\paragraph{Cryogenic optical trapping setup. }
Our detailed experimental setup is shown in Fig.~\ref{fig:setup_detailed}. We optically trap the nanoparticles inside a vacuum chamber connected to a closed-cycle cryostat (attoDRY800 from attocube, nominal cold-plate temperature $4$~K) to lower both the temperature and the pressure of the gas around the particle, thus reducing the fluctuating force disturbing its motion.
The optical tweezers are formed by focusing linearly polarized telecom light with a wavelength $\lambda = 1550$~nm and a power of $1.2$~W.
We use an asymmetric lens system, with a $0.75$~NA trapping lens (Lightpath 355617), and a $0.6$~NA collection lens (Lightpath 355330) to collimate the beam after the trap.
Due to the NA mismatch, about 25\% of the light does not exit the trapping volume and is at least partly absorbed by the cryostat, increasing the temperature of the volume around the trap.
The lenses are encased in a threaded steel mount, and screwed into a threaded holder machined out of electrically insulating polyether ether ketone (PEEK). When performing linear feedback cooling, we apply the voltage needed to drive the particle motion directly to the lenses' mounts.
The PEEK holder is mounted on top of a solid copper post in thermal contact with the cold plate of the cryostat.
The vacuum chamber connected to the cryostat contains two concentric metallic cylinders. Their purpose is to shield the innermost trapping volume of the chamber from hot gas particles in thermal equilibrium with the vacuum chamber at room temperature.
The inner shield, which is made of oxygen-free copper, contains the trapping assembly and is in thermal contact with the cold plate of the cryostat (nominal temperature $4$~K).
The outer cylinder made of aluminium is thermally connected to the middle stage of the cryostat, with a nominal temperature of $40~$K.

\paragraph{Monitoring the temperature.}
We monitor the temperature both at the cold plate and at the PEEK lens holder, and read respectively $6$~K and $57$~K when the laser is on. This discrepancy with the nominal values is due to the heat generated by the absorbed laser power and the low thermal conductivity of the PEEK holder. When the laser is switched off, the trap temperature drops by more than $20$~K in 1 hour. 
The heating due to laser absorption can be remedied by using lenses with equal NA for trapping and collimating the laser, together with optimizing the thermal conductivity of the lens holder.

\paragraph{Measuring the pressure.}
We use a Bayard-Alpert/Pirani combination gauge. Once the base temperature is attained, the gauge reads a pressure of $3 \times 10^{-9}$~mbar at the vacuum chamber thermalized at $295$~K. This is an upper bound for the pressure at the particle's location, which we expect to be orders of magnitude lower~\cite{Micke2019}.

\paragraph{Optical detection setup.}
We use four photodetectors to characterize, stabilize, and localize the  particle in the optical trap.
First, in the forward direction, we make use of a quadrant photodetector (QPD)~(Thorlabs PDQ30C) and a polarisation sensitive libration detector~(homemade balanced detector). We exploit their signals in the characterisation procedure of the particle as detailed below.
Second, we do homodyne and heterodyne detection on the field scattered by the particle back into the trapping lens. We employ a combination of  Faraday rotator and polarizing beamsplitter to deflect the backscattered field from the forward direction.
We derive our feedback signal for cold damping of the particle motion from a balanced, homodyne detector (Thorlabs PDB210C), for which the backscattered light is mixed with a local oscillator (LO) beam whose phase we control with a piezo mirror.

To maximize the detection efficiency, it is essential to properly overlap the signal beam (which has a dipolar scattering pattern collimated by the trapping lens) and the local oscillator, which has a Gaussian mode shape. To this end, we adjust the beam size of the local oscillator with a telescope and carefully tune the propagation distance of signal and reference beam to the detector.

To perform the out-of-loop analysis and sideband thermometry, we use a fiber-coupled balanced heterodyne detector (Newport 2117-FC-M). Here, the LO beam is frequency shifted using two acousto-optic modulators (Gooch\&Housego 3080-1912). The first AOM downshifts the laser frequency by $80~$MHz, while the second upshifts it by $81~$MHz ($79~$MHz) to blueshift (redshift) the LO by $\Og_\text{rf}/(2\pi)=1$~MHz. The resulting detuned LO beam is mixed with the signal in a 50:50 fiber coupler.

\subsection{Particle characterization}\label{sec:particle_characterization}
We optically trap a silica nanoparticle with a nominal diameter of 100~nm (Nanocomposix).
The nanoparticles are provided in aqueous solution which we further dilute in isopropanol and load into the optical trap with a nebulizer.
In order to ensure that the trapped particles are single spherical nanoparticles without rotational degrees of freedom,
we perform a characterization of each object after the trapping process.
In the following, we highlight the two procedures we use to characterize the size and shape of the trapped objects.

\paragraph{Damping rates of transverse motion.}
The first method consists of comparing the damping rate of the transverse $x$ and $y$ modes of oscillation.
At a pressure of few mbar and room temperature, we record a time trace of the $x$ and $y$ oscillation modes on our QPD placed in forward detection (see Fig.~\ref{fig:setup_detailed}). 
Next, we estimate the PSDs from the time traces and fit them to a Lorentzian model.
From the fit we extract the linewidths, and thus the damping rates, of the corresponding modes. Spherical objects have equal damping rates along both axes~\cite{Ahn2018}. Hence, we compute the ratio between the extracted damping rates and use it to identify spherical particles.
Additionally, we estimate the size of the particle using the measured (absolute) damping rate at known pressure and temperature~\cite{Hebestreit2018}.
For the particle used throughout our experiment, we perform this characterization at different pressures ranging from $4$~mbar to $8$~mbar.
We estimate a  diameter of $(106\pm5)$~nm, and a ratio of the damping rate of $0.98\pm0.04$, where the center values are averages among several repetitions of the measurement, whereas the errors are the uncertainties associated to a single measurement (larger than the spread among the measurements).

\paragraph{Libration motion.}
A second characterization method is the detection of a libration motion of the trapped object.
In a linearly polarized electromagnetic field, an anisotropic scatterer aligns itself to the polarization axis and oscillates around this equilibrium position. This libration motion is encoded in fluctuations of the polarization of the scattered light, which we measure using our polarization sensitive balanced photodetector in the forward direction~\cite{Laan2020}. 
If the scatterer is anisotropic, a libration mode is visible at frequencies between 400 kHz and 700 kHz.

\subsection{Parametric particle stabilization}
Throughout our experiment, we stabilize the particle's position along all three axes using parametric feedback cooling~\cite{Gieseler2012}. This reduction of the thermal motion decouples the three center-of-mass degrees of freedom, leading to a three-dimensional, effectively harmonic trapping configuration with the eigenfrequencies $\Ox$, $\Oy$, and $\Oz$ as described in the main text. We emphasize that the parametric feedback cooling along the $z$-axis is much weaker than the linear feedback cooling described in the main text and can hence safely be ignored in the analysis.

We implement parametric feedback cooling using three phase-locked loops (PLL), integrated in a lock-in amplifier (MFLI from Zurich Instruments). Each PLL generates an oscillating signal with constant amplitude and a fixed phase relation to the particle motion along one direction ($x$, $y$, or $z$). We feed the sum of all signals (oscillating at $\Ox$, $\Oy$, and $\Oz$) to a digital squaring unit (STEMLab Red Pitaya), which effectively doubles the frequencies, and use this signal to modulate the intensity of the laser beam using an electro-optic modulator, thereby implementing `PLL-based feedback cooling'~\cite{Jain2016}. We note that on top of the signals at twice the oscillation frequencies, our squarer also generates all sum and difference frequencies between the axes. These spurious signals do not affect the particle's motion in practice since they are off resonant.

\subsection{Data acquisition and postprocessing}
\paragraph{Data acquisition.}
We acquire both the homodyne and heterodyne detector signals by demodulating them at our frequencies of interest using lock-in amplifiers (MFLI from Zurich Instruments).
In particular, we demodulate our homodyne signal close to the eigenfrequency $\Oz$ of the particle and denote the demodulated, complex-valued time trace by $i_\text{hom}[t]$. The square brackets indicate the discrete nature of the time trace, which is an array stored on a computer.
We furthermore demodulate our heterodyne signal close to the two sidebands generated by the particle's motion around the LO frequency (acquired time traces $i_r[t]$ at $\Og_\text{rf} - \Oz$ and $i_b[t]$ at $\Og_\text{rf} + \Oz$), and at the LO frequency itself (acquired time trace $i_\text{car}[t]$ at $\Og_\text{rf}$).
We use $8$th order demodulation filters  with a $3$~dB low-pass frequency of $5$~kHz and a sample frequency of $53.57$~kHz.
For a typical experiment, we acquire 100-second-long demodulated time traces. In addition, we also acquire the homodyne detector signal at baseband ($i_\text{dc}[t]$), which we use both for locking our interferometer with a PI loop integrated into the MFLI and to aid in the postprocessing of the data, as described below.

\paragraph{Postselecting the data.}
\label{sec:dataPost-process}
\begin{figure}[tb]
\includegraphics[width=\textwidth]{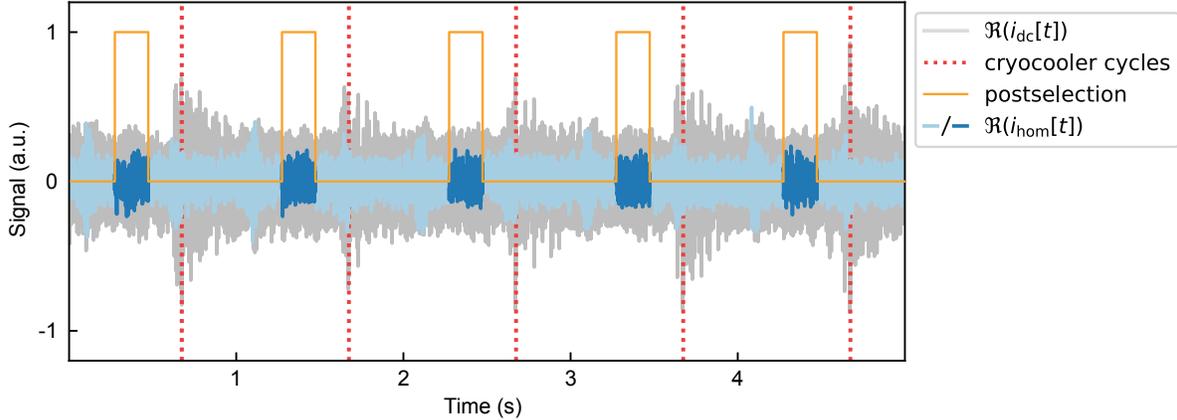}
\caption{{\bf Postselecting the data.}
The compression cycles of the cryocooler are visible in our interferometric signal at baseband ($i_\text{dc}[t]$ in grey). We identify the cycles (red dotted lines) and postselect $300~$ms long intervals (indicator function in orange) of the time traces containing the particle motion (exemplary for $i_\text{hom}[t]$ in blue).}
\label{fig:mask}
\end{figure}

The cryocooler periodically (1 Hz) compresses and expands the helium gas in the cold head, generating periodic mechanical vibrations on the optical table and the trap itself.
These vibrations disturb both the interferometric read-out of the particle's position and its motion.
In our recorded measurements, we hence postselect the time intervals in between the compression cycles.
In Fig.~\ref{fig:mask} we show an example of the homodyne detector signal at baseband ($i_\text{dc}[t]$, grey).
We also show the real part of the particle's signal $i_\text{hom}[t]$ (blue).
We identify the helium compression cycles from $i_\text{dc}[t]$ as burst signals with a repetition period of $1~$s (marked as red dotted lines in Fig.~\ref{fig:mask}).
Finally, we postselect our demodulated time traces ($i_\text{hom}[t]$, $i_r[t]$, $i_b[t]$, and $i_\text{car}[t]$) by choosing 300-ms-long intervals at a fixed delay in between the bursts (indicated by the orange indicator function).
We note that the interval length of $300$~ms is much longer than any time scale of the particle motion.

\paragraph{Estimation of spectral densities.}

After the described postselection, we compute the power spectral densities (PSDs) of the measured time traces. 
We estimate the PSD of the acquired homodyne data according to
\begin{eqnarray}
\Sbar_\text{hom}[\Og] = \langle |i_\text{hom}[\Og]|^2  \rangle,
\end{eqnarray}
where $i_\text{hom}[\Og]$ is the discrete Fourier transform (DFT) multiplied by $\sqrt{T}$ (with the total acquisition time of each realization $T$) of the postselected time traces and  $\langle\dots\rangle$ is the ensemble average over the different realizations.

In contrast to homodyne detection, the heterodyne detector's arm lengths are not actively stabilized, and we have to correct for phase drifts in postprocessing.
These phase drifts are reflected in the phase of the demodulated carrier frequency $i_\text{car}[t]$. Since the frequency components of both motional sidebands have a definite phase relative to the carrier, we can remove the drifts from the time traces by redefining $i_{j}[t] \to i_j[t]e^{-\imu\,\text{arg}( i_\text{car}[t])}$, where $j=r,b$.
After this phase correction, we estimate the PSDs of each sideband as well as the cross-PSD between them as 
\begin{eqnarray}
\Sbar_{rr}[\Og] &=& \langle |i_r[-\Og]|^2\rangle,\\
\Sbar_{bb}[\Og] &=& \langle |i_b[\Og]|^2\rangle,\\
\Sbar_{rb}[\Og] &=& \langle i_r[-\Og] i_b[\Og]\rangle.
\end{eqnarray}
We note that the phase correction described above only affects the cross-PSD $\Sbar_{rb}[\Og]$.

\subsection{Electronic filter characterization}\label{sec:electronic-filter}
\begin{figure}[tb]
\includegraphics{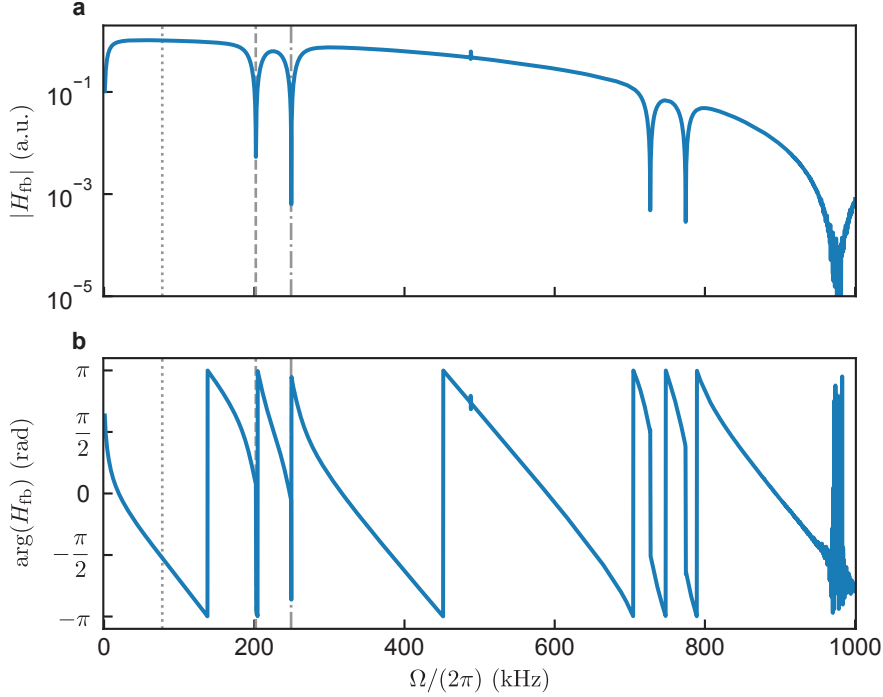}
\caption{{\bf Transfer function of the electronic feedback chain.} {\bf a, b,} Measured magnitude ({\bf a}) and phase ({\bf b}) response of the experimentally used delay filter. The dotted, dashed, and dot-dashed vertical lines mark the location of the resonance frequency of motion along the $z$, $x$, and $y$ axes, respectively.
}
\label{fig:SI-transfer-function}
\end{figure}
In order to model the in-loop dynamics, we need to characterize the transfer function $H_\text{fb}$ of the electronic feedback loop (see Fig.~\ref{fig:setup_detailed}). 
To do so, we perform a network-analyzer measurement of the electronic components in the loop. 
The resulting transfer function is shown in Fig.~\ref{fig:SI-transfer-function}a (absolute value) and Fig.~\ref{fig:SI-transfer-function}b (phase).
Our designed filter contains several elements. First, we have a first-order high-pass filter with a $9$~kHz cut-off frequency, which we use to remove any DC component to prevent saturation of the electronics. Second, we implement two notch filters at $\Omega_x/(2\pi)\approx200$~kHz and $\Omega_y/(2\pi)\approx250$~kHz with a quality factor of $5$.
This way we prevent the feedback from heating the transverse mechanical modes.
We also observe two copies of such filters at around $750$~kHz. This is due to aliasiang of the signal
during the frequency sweep measurements. In fact, the sampling rate is at $\approx977$~kHz, resulting in a Nyquist frequency of $488.5$~kHz.
Finally, we introduce a time delay such that at $\Og_z$ the phase response is $-\pi/2$. Supposing that the phase contribution of the high-pass and the notch filters are negligible at $\Og_z$, one can tune the delay time such that $\Og_z \tau = \pi/2 + 2\pi n$, where $n$ is an integer. For any $n>1$, the larger phase slope lowers the value of the feedback gain at which the closed-loop system  becomes unstable, limiting the cooling performance. Therefore, we choose to implement the smallest possible time delay, which in our case is $\tau\approx 3.2\,\mu\text{s}$.

\subsection{Detection noise characterization}

\begin{figure}[hb]
\includegraphics{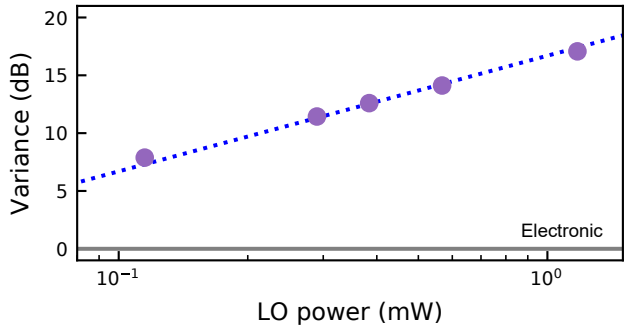}
\caption{{\bf Detection noise characterization.}
Variance of the laser noise as a function of local oscillator power in homodyne detection.
The variance, expressed in dB, is normalized to the variance of the electronic noise floor of the detector (grey). The dotted blue line provides a guide for the eyes for the linear dependence between variance and power of the beam.
}
\label{fig:det_characterization}
\end{figure}

For feedback-based ground-state cooling, it is critical that our in-loop, homodyne detection noise is limited by the shot noise of the optical field.
In Fig.~\ref{fig:det_characterization} we show the measured noise power on the homodyne detector when only the LO beam is switched on (and the particle signal is blocked) as a function of the LO power. The noise power is obtained by integrating the measured PSD from $60$ to $90$~kHz and normalizing it by the detector electronic background-noise power (indicated by the grey line). The LO power is tuned by rotating a half-wave plate in front of a polarizer. We observe that the noise power increases linearly with the LO power, thus indicating that our detection is shot-noise limited. In the experiment, we operate at $560~\mu$W of LO power, where the optical shot noise is $14$~dB above the electronic noise floor. %

\section{Out-of-loop heterodyne measurements}\label{sec:out-of-loop-analysis}
In this section, we detail the characterization of and analysis performed on the out-of-loop heterodyne measurements.

\subsection{Sideband-asymmetry thermometry}
We record the PSDs of the two mechanical sidebands around the heterodyne local oscillator. The two sidebands differ in their carried noise power. This asymmetry is related to the mechanical zero-point fluctuations and can be used to extract the phonon occupation \cite{Underwood2015, Tebbenjohanns2020}.
\begin{figure}[tb]
\includegraphics[width=\textwidth]{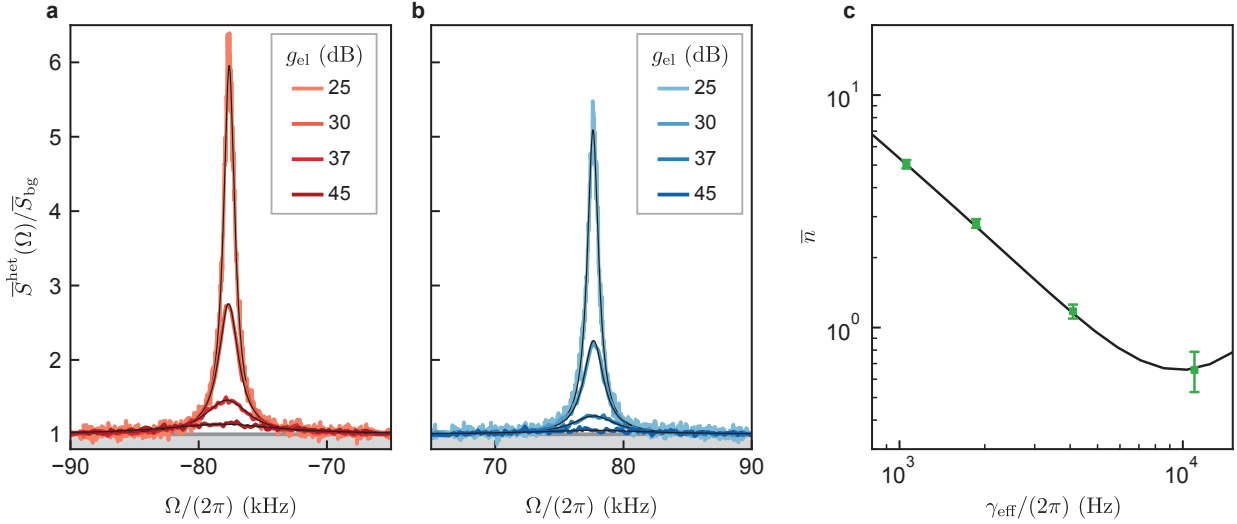}
\caption{{\bf Sideband asymmetry in out-of-loop heterodyne measurements.}
{\bf a, b,} Stokes ({\bf a}) and anti-Stokes ({\bf b}) sidebands, at different electronic feedback gains, normalized to the estimated background level (grey line). Each sideband pair is simultaneously fitted to a theoretical model.
{\bf b,} Mechanical occupations (green squares) at different feedback gains. The black solid line is a theoretical model based on an ideal delay filter with parameters estimated from the in-loop spectra.
The errorbars are obtained by propagating the fit uncertainties (s.d.) of the areas.}
\label{fig:SI-hete-calib}
\end{figure}
In Fig.~\ref{fig:SI-hete-calib}a,b, we show the PSDs of the two sidebands for different feedback gains. 
In order to quantitatively assess the mechanical energy of the particle, we extract the area underneath each sideband.
We fit each pair of sidebands simultaneously to a theoretical model $\Sbar_\text{jj}(\Og) = \Sbar_\text{bg}^j + |\chi_\text{eff}(\Og)|^2 \Sbar_{FF}^j$, where $j=r,b$ and the mechanical susceptibility takes the form $\chi_\text{eff}(\Og)=m^{-1}/(\Oz^2-\Og^2-\imu\Og\Geff)$. In the fit model, we allow the two sidebands to assume different force noise $ \Sbar_{FF}^j$,  and background $\Sbar_\text{bg}^j$ values, but we constrain them to have the same resonance frequency $\Oz$, and linewidth $\Geff$.

The fitted force-noise values are a direct measure of the enclosed area in the two sidebands. Thus, the occupation $\nbar$ can be extracted according to
\begin{eqnarray}
\frac{\Sbar_{FF}^r}{\Sbar_{FF}^b} = 1+\frac{1}{\nbar}.
\end{eqnarray}
The uncertainties of these areas crucially depend on the precision of the background-noise estimation from the fitting routine, especially at the largest feedback gain where the signal-to-noise ratio becomes small. The estimation of the occupation using the spectral cross-correlation, detailed in the following subsection, is robust against this possible source of error.

Another possible source of systematic error is a frequency-dependent response of the acquisition chain (photodetector and DAQ). To rule out this effect, we measure the motional sidebands both using a positive and a negative frequency for the heterodyne local oscillator $\omega_\text{LO} =\omega_L - \Og_\text{rf}$, where $\omega_L$ is the frequency of the laser and $\Og_\text{rf}/(2\pi)=\pm1$~MHz denotes the frequency shift induced with the AOMs \cite{Tebbenjohanns2020}. We then extract the phonon occupation according to
\begin{eqnarray}\label{eq:phonon-occ-both}
\overline{n} = \left(\sqrt{\frac{S_{FF}^{r,+}S_{FF}^{b,-}}{S_{FF}^{b,+}S_{FF}^{r,-}}}-1\right)^{-1},
\end{eqnarray}
where the $^\pm$ superscripts stand for the sign of the LO frequency shift.
Using this method, any frequency dependence of the transfer function of the measurement chain is cancelled.
In Fig.~\ref{fig:SI-hete-calib}b, we show as green squares the phonon occupations estimated from the asymmetry of the measured heterodyne spectra. We also show as a black line the theoretical cooling model extracted from the in-loop analysis (see Sec.~\ref{sec:in-loop-SI}). The errorbars are obtained by propagating in Eq.~\eqref{eq:phonon-occ-both} the fit uncertainties (s.d.) of the four areas extracted from the fits (two areas per each frequency of the local oscillator). The larger errorbars for lower occupations reflect the reduced signal-to-noise ratio in the PSDs.

\subsection{Cross-correlation thermometry}

\begin{figure}[tb]
\includegraphics[width=\textwidth]{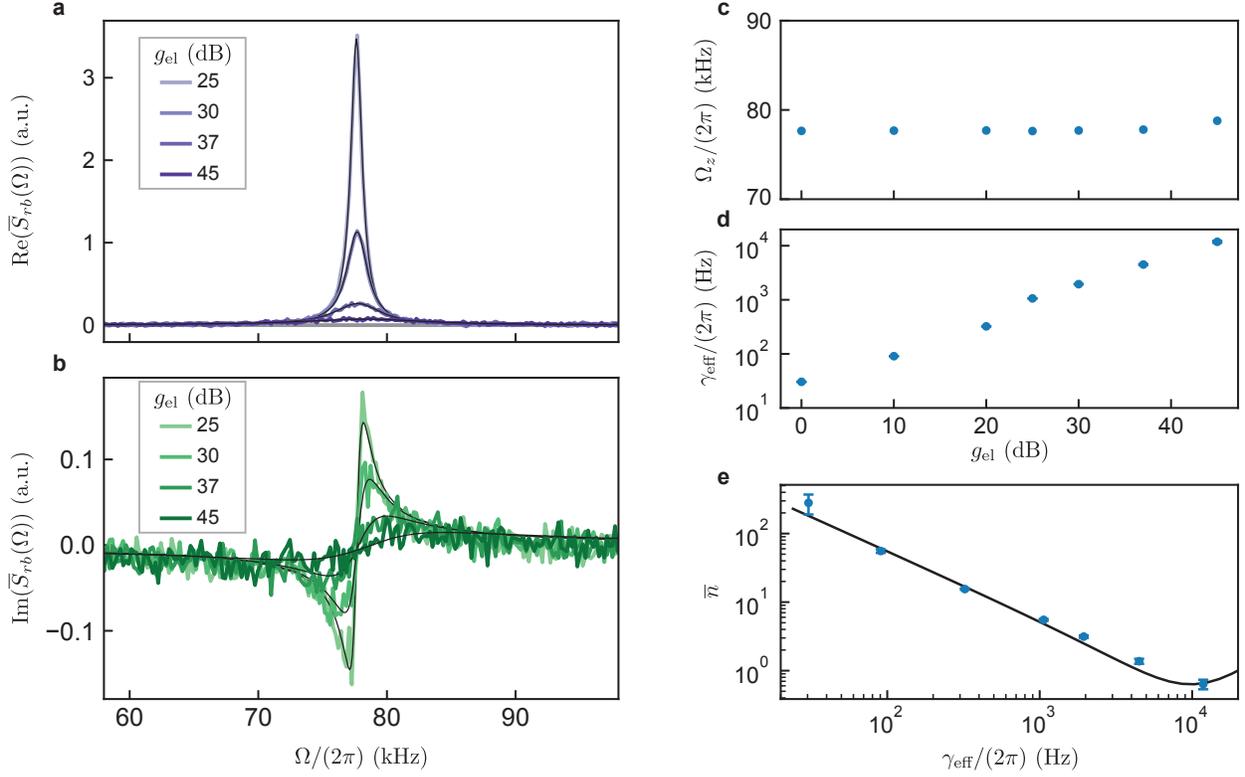}
\caption{{\bf Sidebands cross-correlations in out-of-loop heterodyne measurements.}
{\bf a, b,} Real ({\bf a}) and imaginary ({\bf b}) parts of cross-spectra, at different electronic feedback gains. Each pair is simultaneously fitted to a theoretical model and the results are shown as black lines. The grey line marks the zero as a reference. 
{\bf c, d,} Fitted mechanical resonance frequency ({\bf c}) and effective linewidth ({\bf d}) at different electronic gains.
{\bf e, } Extracted mechanical occupations as a function of fitted effective linewidths. The black line is a theoretical model based on an ideal delay filter and on parameters estimated from the in-loop spectra.
The errorbars are obtained by the fit uncertainties (s.d.).}
\label{fig:SI-cross-corr}
\end{figure}

To corroborate the measured occupations from the asymmetry of the motional sidebands, we perform an additional thermometry measurement based on the quantum correlations between the Stokes and anti-Stokes sidebands \cite{Purdy2017, Shkarin2019}.
The cross-PSD between these two sidebands can be expressed as
~\cite{Purdy2017}
\begin{equation}
    \Sbar_{rb}(\Og) \propto \Sbar_{zz}(\Og) + \frac{\imu\hbar}{2\pi} \RE{\chi_\text{eff}(\Og)},
\end{equation}
which we simplify to 
(see also Supplement of~\cite{Shkarin2019} for a derivation with $\Geff\ll\Oz$) 
\begin{equation}\label{eq:cross-spectrum}
\begin{split}
\Sbar_{rb}(\Og)  ~=~& R |\chi_\text{eff}(\Og)|^2\left(\nbar+\frac{1}{2} + \imu \frac{\Og^2-\Oz^2}{2\Oz\Geff}\right) 
~\xrightarrow{\Geff\ll\Oz}~  R |\chi_\text{eff}(\Og)|^2\left(\nbar+\frac{1}{2} + \imu \frac{\Og-\Oz}{\Geff}\right),
\end{split}
\end{equation}
where $R$ is a constant proportionality factor. In particular, the imaginary part of Eq.~\eqref{eq:cross-spectrum} arises from correlations induced by the zero-point fluctuations, and solely depends on spectroscopic quantities easily accessible (resonance frequency $\Oz$ and linewidth $\Geff$) and {\it not} on the occupation $\nbar$. Therefore, one can use the imaginary part of the correlator in Eq.~\eqref{eq:cross-spectrum} as a calibration for the real part in Eq.~\eqref{eq:cross-spectrum}, which directly yields the phonon occupation.

Equation~\eqref{eq:cross-spectrum} assumes that the reference frame in which such cross-correlations are computed is only defined by the reference local oscillator.
In practice, the measured cross-PSD $\tilde S_{rb}(\Omega) = e^{2\imu\theta} \Sbar_{rb}(\Omega)$ is rotated by an angle $\theta$, which is the heterodyne LO angle with respect to the signal at the time when the data acquisition starts. Due to drifts of the interferometer arm lengths, the value of $\theta$ drifts at a slow rate.
In order to factor this out, we exert a coherent, off-resonant driving force on the particle at $90$~kHz. 
In the ideal reference frame, the spectral component at the frequency of this coherent drive is purely real. 
Thus, we can extract $\theta$ from the phase of the measured correlator at $90$~kHz according from the expression $2\theta = \text{arg(}\tilde S_\text{rb}[2\pi\times90~\text{kHz}]\text{)}$, in order to then rotate the measured cross-PSD into the ideal reference frame.
After this calibration, we  simultaneously fit the real and imaginary parts of the measured cross-spectra to Eq.~\eqref{eq:cross-spectrum}. We choose as free parameters the mechanical resonance frequency $\Oz$, the linewidth $\Geff$, and the overall scaling factor for the real part and imaginary parts, respectively $c_r\equiv R\left(\nbar+1/2\right)$ and $c_i\equiv R$. Finally, we compute the occupation from the ratio of the two scaling factors, that is, $\overline{n} = c_r/c_i - 1/2$.

In Fig.~\ref{fig:SI-cross-corr} we show examples of measured and fitted spectra, as well as an overview of the fitted parameters and extracted occupations.
We stress that, in contrast to the sideband asymmetry thermometry (detailed in the previous subsection), the method presented here does not rely on the precise subtraction of a background noise to estimate the phonon occupation. This makes the method detailed here more robust against experimental drifts.

\section{In-loop homodyne measurements}\label{sec:in-loop-SI}
Homodyne-based feedback control of mechanical motion has been extensively studied both theoretically \cite{Mancini1998, Doherty1999b, Genes2008} and experimentally \cite{Cohadon1999, Poggio2007, Wilson2015, Rossi2018}.
Here, we summarize the main equations used in our analysis of the in-loop measured spectra, and we report the experimental characterization and methods employed.

\subsection{In-loop detection theory}\label{sec:theory-feedback}
We model the dynamics of the feedback-controlled quantum system with quantum Langevin equations \cite{Genes2008}. This framework allows dealing with the non-Markovianity associated with any realistic feedback loop, which renders  the adoption of a standard Lindblad master equation approach impossible \cite{Wiseman2010}.
The system under control is a levitated particle in an initial thermal state, undergoing a linearized optomechanical interaction with the trapping field \cite{Rodenburg2016}. The initial state is therefore a Gaussian one and the linear dynamics of both the evolution and measurement preserves the Gaussian nature of the states over time.
Thus, the quantum dynamics can be described in terms of an analogous classical system, with the additional constraints of (i) zero-point fluctuations present in both the optical and the mechanical degrees of freedom, and (ii) the non-zero bound of the Heisenberg measurement-disturbance relation \cite{Wiseman2010}.

Here, we are interested in modelling the spectra measured by the in-loop homodyne detector, $\Sbar_{zz}^\mathrm{hom}$, as well as the actual displacement spectra, $\Sbar_{zz}$.
In order to stabilize our system, we never release it completely from feedback. For the smallest gain setting $g_\text{el}=0$~dB in the main text, the induced damping rate $\gamma_m$ largely exceeds the intrinsic damping rate given by the bath interaction but is small enough that we can neglect any in-loop effects (i.e. the resonant spectral response is much larger than the imprecision).
The measured homodyne spectrum is then
\begin{eqnarray}\label{eq:homo-spectrum-no-feedback}
\Sbar_{zz}^\mathrm{hom}(\Og) &=& \Simp + |\chi_m(\Og)|^2 \Sfftot,
\end{eqnarray}
where $\chi_m(\Og) = m^{-1}/(\Oz^2-\Og^2-\imu \gamma_m \Og)$ is the mechanical susceptibility, and $\Simp$ and $\Sfftot$ are, respectively, the imprecision and the total force noise.

The total force noise contains all the fluctuating forces acting on the resonator, which we write as
\begin{eqnarray}
\Sfftot = \frac{\hbar^2}{2\pi \zzpf^2} \Gtot
\end{eqnarray}
where the total decoherence rate $\Gtot=\Gexc + \Gqba$, contains the decoherence rate associated with the quantum backaction due to random photon recoils $\Gqba$, and the excess decoherence rate $\Gexc$, comprising all other sources of decoherence such as collisions with gas molecules and heating due to technical laser noise.
The imprecision noise, $\Simp$, represents the background floor of the measured spectrum, and can be written as
\begin{eqnarray}
\Simp = \frac{\zzpf^2}{8\pi\Gmeas},
\end{eqnarray}
where $\Gmeas=\eta_d\Gqba$ is the measurement rate and $\eta_d$ the total detection efficiency.

We use the measured homodyne photocurrent as the input signal on the control feedback loop, the complex transfer function of which we call $H_\text{fb}(\Og)$.
In general, the only restrictions to this transfer function for an experimentally viable loop are (i) to be causal, for real-time control, and (ii) to maintain the controlled system stable, for continuous operation \cite{Wiseman2010}.
A possible causal filter satisfying the stability requirement (up to a certain gain) is a pure delay filter, whose transfer function is
\begin{eqnarray}\label{eq:ideal_delay_filter}
H_\text{fb}(\Og) = m \Og_z \gamma_\text{fb} e^{\imu \Og\tau},
\end{eqnarray}
where $\gamma_\text{fb}$ is the feedback gain in units of angular frequency and $\tau$ the chosen time delay. 
Given this freedom, the transfer function can in principle be optimized in order to minimize a cost function of the controlled system's degrees of freedom, achieving optimal control \cite{Wieczorek2015}. For example, in the case of ground state preparation, such a cost function is represented by the mechanical energy, which is a quadratic function of the mechanical degrees of freedom for a harmonic oscillator. This property, combined with the linear dynamics and the involved Gaussian state, allows us to directly make use of the known results from classical linear-Gaussian-quadratic (LGQ) control theory \cite{Garbini1996}.
In practice when dealing with a high-Q mechanical resonator, using a non-optimal filter results in slightly worse performances, accompanied by a great simplification of the experimental implementation.
Thus, we decide to follow the latter strategy. We experimentally implement a digital delay filter (see Sec.~\ref{sec:electronic-filter}),
which is  the optimal filter for minimizing the mechanical energy only in the limit of a strongly underdamped oscillator.

Once we close the loop, Eq.~\eqref{eq:homo-spectrum-no-feedback} is not valid anymore, and the in-loop effects should be properly included to interpret the homodyne measurements.
By following a standard derivation \cite{Wilson2015, Rossi2018}, we arrive at the following expression for the in-loop homodyne spectrum
\begin{equation}\label{eq:in-loop-spectrum}
\begin{split}
\Sbar_{zz}^\mathrm{hom}(\Og) =&~|\chi_\text{fb}(\Og)|^2\left(\Sfftot + |\chi_m(\Og)|^{-2}\Simp\right) \\
\xrightarrow{\gamma_\text{fb}\gg\gamma_m}&~ \Simp + ~|\chi_\text{fb}(\Og)|^2\left(\Sfftot -m^2 \gamma_\text{fb}^2 \Og^2 \Simp\right),
\end{split}
\end{equation}
where $\chi_\text{fb}(\Og)^{-1} = \chi_m(\Og)^{-1}-H_\text{fb}(\Og)$ is the mechanical susceptibility modified by the feedback loop. In particular, the last term on the right-hand side of Eq.~\eqref{eq:in-loop-spectrum} describes the induced correlation between the imprecision noise and the mechanical displacement, which is driven by the same imprecision noise via the feedback loop. For large gains such that $m\Og_z\Gfb \gtrsim \sqrt{\Sfftot/\Simp}$, these correlations result in measured spectral values lower than the imprecision noise, an effect known as {\it noise squashing} \cite{Cohadon1999, Aspelmeyer2014}.

Due to the presence of such correlations, Eq.~\eqref{eq:in-loop-spectrum} does not describe the actual mechanical displacement spectrum at large gains, thus we cannot estimate the occupation from its integration. Rather, we need to calculate the actual position spectrum under feedback control as~\cite{Wilson2015, Rossi2018}
\begin{eqnarray}\label{eq:actual-position-spectrum}
\Sbar_{zz}(\Og)&=&|\chi_\text{fb}(\Og)|^2\left(\Sfftot + |H_\text{fb}(\Og)|^{2}\Simp\right).
\end{eqnarray}

\subsection{In-loop homodyne thermometry}
\begin{figure}[tb]
\includegraphics{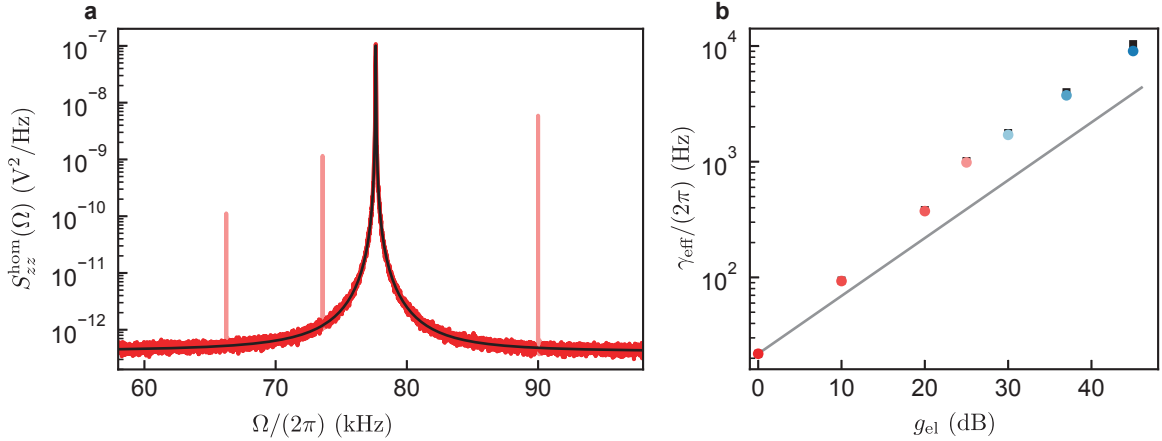}
\caption{{\bf Fit results.}
{\bf a,}  Reference displacement spectrum measured by the homodyne detector at the smallest feedback gain, with a fit to a model (black line). In light red we show the spectral features excluded from the fits.
{\bf b,} Fitted feedback gains, $\Geff$, as a function of the experimentally tunable electronic gains $g_\text{el}$. Coloured dots come from fitting the corresponding spectra shown in Fig.~3a of the main text. The black squares are the full-width-half-maximum extracted from the computed actual displacement spectra. The grey line is a guide for the eye, and represents the expected linear relation.}
\label{fig:SI-inloop}
\end{figure}

In the following section, we describe the fitting procedure employed for the in-loop spectra and for the extraction of the phonon occupation, based on the theory outlined in Sec.~\ref{sec:theory-feedback}.

First, we analyze the initial spectrum, shown in Fig.~\ref{fig:SI-inloop}a, which we take as a reference. This reference spectrum results from a combination of linear and PLL-based parametric feedback cooling applied to the particle, necessary to keep it trapped in ultra-high vacuum. 
In this configuration, the feedback gain is kept low such that the spectral value at the mechanical resonance frequency, $\Sbar^\text{hom}_{zz}(\Og_z)$, is much larger than the imprecision noise, $\Simp$. Assuming a time delay of $\tau=\pi/(2\Og_z)$ and approximating the delay filter's phase response constant around the mechanical resonance frequency (which is valid in the limit $\Gm\tau\ll1$), the feedback only modifies the linewidth of the mechanical susceptibility in Eq.~\eqref{eq:homo-spectrum-no-feedback}, with negligible induced correlations.
The effective linewidth can be expressed as $\Geff \approx\Gfb=\Gm g_\text{el}$, where $\Gm$ is the induced linewidth at unity gain $g_\text{el}=1$.

We fit the initial reference spectrum to the model of Eq.~\eqref{eq:homo-spectrum-no-feedback}. From the fit, we extract a mechanical resonance frequency of $\Oz/(2\pi)=77.6$~kHz and a linewidth of $\Gm/(2\pi)=21.9$~Hz. In addition we extract the total force noise and the imprecision noise, which are still in electrical units at this stage.

Next, we record homodyne spectra as we increase the linear feedback gain.
The spectra are shown in Fig.~3a of the main text. We fit each of these spectra to the full in-loop model of Eq.~\eqref{eq:in-loop-spectrum}. We fix the mechanical resonance frequency, linewidth, total force noise and imprecision noise from the previous fit of the initial spectrum. Also, we independently measure the feedback loop's complex transfer function and interpolate it (see Sec.~\ref{sec:electronic-filter}), then we use it in the definition of the fitting function. The only free parameter left is the feedback gain ($\Gfb$) multiplying the entire transfer function. In Fig.~\ref{fig:SI-inloop}b we show the fitted gains for each spectrum as a function of the electronic gain, showing the expected linear relationship.

We notice that we exclude in our analysis three spectral features, highlighted in light red in Fig.~\ref{fig:SI-inloop}a. The components at $66.3$~kHz is an electronic noise peak generated by the ion pressure gauge used in the experiment. Regarding the component at $73.5$~kHz, we hypothesize it originates in the frequency noise of the laser, and it translates into a force via a residual longitudinal standing wave present in the optical trap (due to spurious back-reflections). Finally, the component at $90$~kHz is a calibration force we employ to analyze the heterodyne signal (see Sec.~\ref{sec:out-of-loop-analysis}). All three components are coherent (i.e. their linewidth is Fourier-limited), thus do not contribute to the total displacement fluctuations.
The same three components are excluded also from the out-of-loop measurements analysis.

After fitting the in-loop spectra, we extract the mechanical energy by integrating the  position and momentum spectra, according to
\begin{eqnarray}\label{eq:mech-occ-integral}
\overline{n} = \frac{m\Og_z^2 \langle z^2 \rangle + \langle p_z^2 \rangle /m}{2\hbar\Og_z}-\frac{1}{2} =\int_0^\infty \diff \Og ~  \left(1+\frac{\Og^2}{\Og_z^2}\right)\frac{\Sbar_{zz}(\Og)}{2\zzpf^2} - \frac{1}{2}.
\end{eqnarray}
We notice that the integration is done over both the position and the {\it momentum} spectrum, as deviations from the equipartition theorem become significant when $\Geff\approx\Oz$ \cite{Genes2008}.
For each electronic gain, we compute the actual displacement spectrum by using Eq.~\eqref{eq:actual-position-spectrum}, with the set of parameters estimated from the initial spectrum ($\Oz$, $\Gm$, $\Sfftot$ and $\Simp$), the fitted gain from the in-loop spectrum ($\Gfb$), and the interpolated transfer function ($H_\text{fb}$).
As a final step, we need to calibrate the measured mechanical energy from units of $V^2$ to phonons.
To do that, we anchor the energy extracted from the in-loop homodyne spectrum at $g_\text{el}=25$~dB to the one extracted by the sideband asymmetry thermometry at the same gain, that is $\overline{n}_{25~\text{dB}}=5.1\pm0.1$.
The calibrated occupations extracted from the in-loop analysis are reported in Fig.~3b of the main text.
The effective linewidth in this figure are extracted as full-width-half-maximum from the computed displacement spectrum, $\Sbar_{zz}(\Og)$.

\bibliography{Literature}